\documentclass[aps,twocolumn,showpacs,superscriptaddress,letterpaper]{revtex4-1}
\usepackage{graphicx,amsmath,amsfonts}
\usepackage{tabularx}
\usepackage{bm}
\usepackage{mathrsfs}
\usepackage{appendix}
\usepackage{bbm}
\usepackage{color}
\def \Svk {S(\vk ,\omega )}
\def \Sab {S_{\alpha \beta }(\vk ,\omega )}

\def \vR {\bf R}

\def \vk {{\bf k}}

\def \vS {{\bf S}}

\def \vz {{\bf z}}
\def \vx {{\bf x}}
\def \vy {{\bf y}}
\def \vr {{\bf r}}

\def \vQ {{\bf Q}}

\begin{document}

\title{Topological and Magnetic Properties of a Non-collinear Spin State on a Honeycomb Lattice in a Magnetic Field}

\author{Randy S. Fishman}
\email{corresponding author:  fishmanrs@ornl.gov
\newline
This manuscript has been authored in part by UT-Battelle, LLC, under contract DE-AC05-00OR22725 with the US Department of Energy (DOE). The US government retains and the publisher, by accepting the article for publication, acknowledges that the US government retains a nonexclusive, paid-up, irrevocable, worldwide license to publish or reproduce the published form of this manuscript, or allow others to do so, for US government purposes. DOE will provide public access to these results of federally sponsored research in accordance with the DOE Public Access Plan (http://energy.gov/downloads/doe-public-access-plan)}

\affiliation{Materials Science and Technology Division, Oak Ridge National Laboratory, Oak Ridge, Tennessee 37831, USA}
\author{Daniel Pajerowski}
\affiliation{Neutron Sciences Division, Oak Ridge National Laboratory, Oak Ridge, Tennessee 37831, USA}

\date{\today}

\begin{abstract}  We study the Berry curvature and Chern number of a non-collinear spin state on a honeycomb lattice that evolves from coplanar to ferromagnetic with a 
magnetic field applied along the $z$ axis.  The coplanar state is stabilized by nearest-neighbor ferromagnetic interactions, single-ion anisotropy along
$z$, and Dzyalloshinskii-Moriya interactions between next-nearest neighbor sites.  Below the critical field $H_c$ that aligns the spins, the 
magnetic unit cell contains $M=6$ sites and the spin dynamics contains six magnon subbands.  
Although the classical energy is degenerate wrt the twist angle $\phi $ between nearest-neighbor spins, the dependence of
the free energy on $\phi $ at low temperatures is dominated by the magnon zero-point energy, which contains extremum at
$\phi =\pi l/3$ for integer $l$.  The only unique ground states GS($\phi )$ have $l=0$ or 1.  
For $H < H_c'$, the zero-point energy has minima at even $l$
and the ground state is GS(0).  
For $H_c' < H < H_c$, the zero-point energy has minima at odd $l$ and 
the ground state is GS($\pi/3$).  In GS(0), the magnon density-of-states exhibits five distinct phases with increasing field
associated with the opening and closing of energy gaps between the two or three magnonic bands, each containing between 1 and 4 four magnon subbands. 
While the Berry curvature vanishes for the 
coplanar $\phi=0$ phase in zero field, the Berry curvature and Chern numbers exhibit signatures of the five phases at nonzero fields below $H_c'$.  
If $\phi \ne \pi l/3$, the Chern numbers of the two or three magnonic bands are non-integer.  
We also evaluate the inelastic neutron-scattering spectrum $S(\vk ,\omega )$ produced by the six magnon subbands in all five phases of GS(0) and in GS($\pi/3$).
Whereas the Berry curvature and Chern number are sensitive to changes in the magnon density-of-states, the inelastic spectrum 
is sensitive to changes in the number and intensity of the magnon subbands rather than their distribution into bands.
These results indicate that special care must be taken when treating the topological properties of complex non-collinear spin states with $M>2$ magnetic sublattices.
		\end{abstract}


\maketitle

\section{Introduction}

Remarkable advances in the field of  ``magnonics" \cite{Wang21, Chumak22, Sheka22} over the past 14 years were initiated by the discoveries 
that magnons or quanta of spin excitations produced the 
thermal Hall \cite{Onose2010, Ideue12, Hirschberger15a, Hirschberger15b, Murakami17} and
Seebeck \cite{Uchida10, Wu16} effects.  Technological applications of magnons exploit the absence of a magnetic charge, which allows magnons to 
travel large distances without causing Joule heating \cite{Buttner2000}.  Consequently, magnons have been proposed to replace electrons in devices that 
involve information storage and communication.   

Other applications have been recently inspired by the topological properties of magnon edge modes.   
Prior to its observation \cite{Onose2010} in the ferromagnetic (FM) insulator Lu$_2$V$_2$O$_7$,
the magnon Hall effect was predicted by Katsura {\it et al.} \cite{Katsura2010} based on a Kubo formula for the 
temperature dependence of the thermal conductivity $\kappa^{xy}(T)$ due to the effective magnetic field produced
by the Berry curvature (BC).  As discussed further below,
the Chern number (CN) can be evaluated by integrating the BC over the first Brillouin zone (BZ).
Due to the bulk-boundary correspondence \cite{Mat11a, Mat11b, Mong11, Zhang13, Mook14a}, 
the CN also gives the number of edge modes when a two-dimensional topological material is cut into a ribbon. 
Therefore, these magnon edge states are ``topologically protected" and 
resistant to decay through impurity scattering.  This topological protection offers many prospective 
technological applications.

However, most studies of topological magnets have been confined to collinear  
FM \cite{Shindou13, Owerre16, Owerre16b, Mook21, Li21b, Lu21, Fishman23} or antiferromagnetic (AF) 
\cite{Zyuzin16, Cheng16, Nakata17, Kawano19, Neumann22, Go24} systems.  Commonly studied non-collinear (NC)
systems \cite{Sticlet12, Fujiwara22,Chen23} contain only $M=2$ magnetic sublattices.  Such 
systems have the advantage that models with two subbands can be mapped
onto a Haldane model \cite{Haldane88}.  After obtaining analytic results for the spin-wave 
(SW) frequencies $\omega_n(\vk )$ and eigenvectors $\vert u_n (\vk )\rangle $, it is then straightforward
to evaluate the BC
\begin{equation}
{\bf \Omega}_n (\vk)=\frac{i}{2 \pi} 
\biggl\{ \frac{\partial }{\partial \vk } \times
\langle u_n(\vk ) \vert \frac{\partial }{\partial \vk }\vert u_n(\vk) \rangle \biggr\}
\label{EqBerry}
\end{equation}
and CN
\begin{equation}
C_n=\int_{\rm BZ} d^2k\, \Omega_{nz}(\vk )
\label{Chern}
\end{equation}
for wavevector $\vk $ and eigenmode $n=1$ or 2, 
where $\vk $ is integrated over the first BZ \cite{Shindou13} of a two-dimensional system normal to $\vz$.

For practical reasons, it is of great importance to understand the topological properties of more complex NC spin states with $M>2$ magnetic sublattices.  
Many materials of technological interest exhibit complex NC spin states even in zero field \cite{Wang23, Martini23, Li23}.   
In addition, a great deal can be learned about 
zero-field FM and AF states from the NC states that develop in a magnetic field.
Of course, the switching or rotation of a collinear state in a magnetic field alway involves an intermediate 
NC state \cite{Soenen23, Hayami23, Fernandes24, Guo24, Jana24}. 

Unfortunately, complex NC spin states pose special challenges.  Due to the large magnetic unit cells with $M>2$ sites, it is usually necessary to 
evaluate the SW frequencies and eigenvectors numerically.  The absence of an analytic result for the eigenvector $\vert u_n (\vk )\rangle $ means 
that its phase can change from one $\vk $ point to another.  Therefore, special care must be taken when numerically evaluating the derivatives 
$\partial \vert u_n(\vk )\rangle /\partial \vk $ that enter the BC.

We now evaluate the BC and CN for a model system based on a honeycomb (HC) lattice with FM nearest-neighbor interaction $J$, single-ion
anisotropy $K$ and magnetic field $H$ along $z$, and a Dzyalloshinskii-Moriya (DM) interaction $D$ due to broken inversion symmetry also along $z$. 
When $\vert D\vert $ is sufficiently large, the DM interaction will force the spins to lie down in the $xy$ plane for zero field.  
The magnetic field will then tilt the spins out of the plane towards the $z$ axis until
they align along $\vz$ at a critical field $H_c$.  Because the magnetic unit cell contains $M=6$ spins, the spin 
dynamics will contain 6 eigenstates (defining subbands $n=1,\ldots,6$) that must be solved numerically.
This is one of the simplest models that preserves the ${\cal C}_3$ symmetry of the HC lattice yet still demonstrates the
topological properties of a non-trivial NC spin system as it evolves with field.  

Surprisingly, even this relatively simple model poses some serious challenges and produces some rather interesting results.  
Below $H_c$, the spin ground state (GS) denoted as GS($\phi $) depends on the twist angle $\phi $ between neighboring spins on the lattice, 
as seen in Figure 1(a).  While the classical energy is degenerate wrt $\phi $, the zero-point energy is minimized when 
$\phi =0$ for $H<H_c'$ and when $\phi =\pi/3$ for $H_c' < H < H_c$.  In GS(0) below $H_c'$, the magnon density-of-states (DOS)
$\rho (\epsilon )$ defines five phases that depend on the structure of the 6 subbands.  Transitions between these five phases occur 
due to the opening and closing of gaps in the magnonic band structure.  The structure of the DOS is reflected in the BC and CN, which are different in all five phases.   
In GS($\pi/3$) above $H_c'$, the DOS contains only two magnonic bands.

For $\phi=0$, the zero-field GS with coplanar spins have no BC out of the plane.  Therefore, the CNs of the two zero-field bands are both zero.  
But in the FM, the two magnon bands have CNs of $\pm 1$.  In order for the total CNs to be preserved across each phase boundary from zero field to $H_c$, 
the system must pass through at least one phase with three or more bands.  The band structure of our model passes from two (phase $i$) to three (phase $ii$) to two 
(phase $iii$) to three (phase $iv$) and then back to two (phase $v$) bands.

We also compare the topological properties of this model with its inelastic neutron-scattering (INS) spectra.  
The INS spectra changes very little across the five phases of GS(0) but then changes rapidly across GS($\pi/3$) as the 
$M=6$ modes collapse onto $M=2$ modes as the FM state is approached when $H\rightarrow H_c$.
Hence, the INS spectrum is sensitive to changes in the 
number and intensity of the magnon subbands rather than their distribution into bands.

\begin{figure}
\begin{center}
\includegraphics[width=8.5cm]{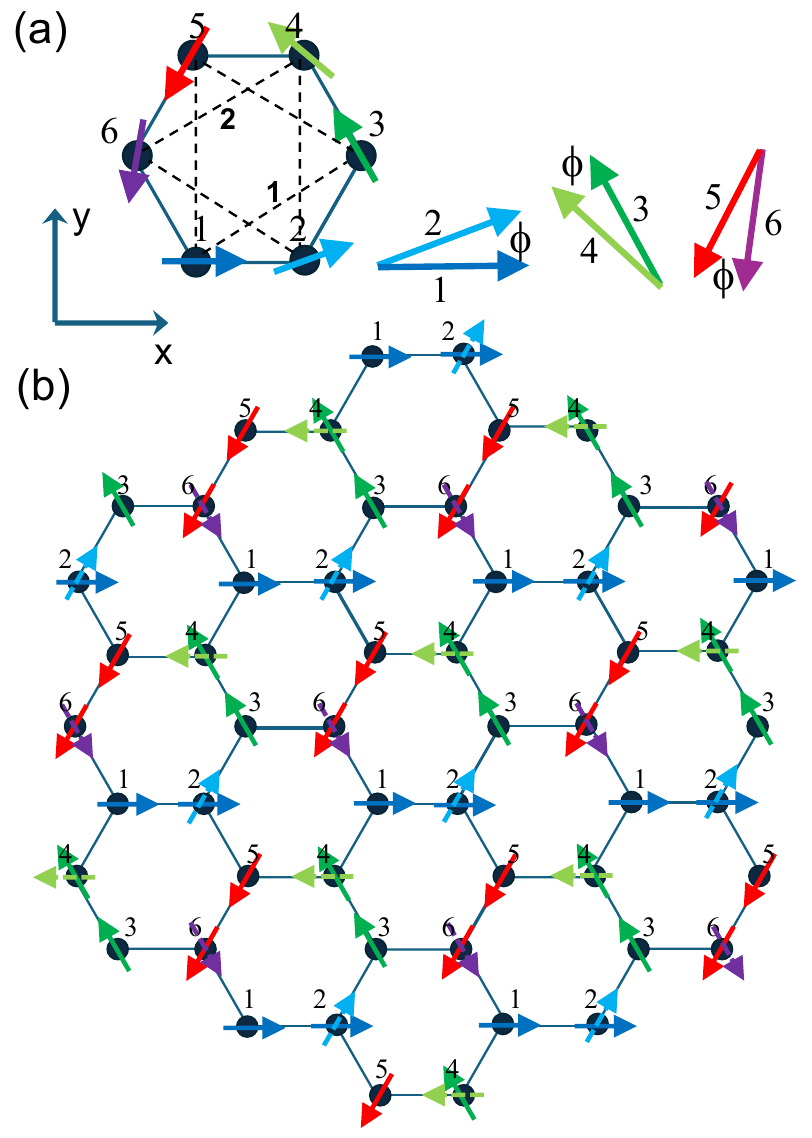} 
\end{center}
\caption{
(a) The twist angle $\phi $ between spins $\{1,2\}$, $\{3,4\}$ and $\{5,6\}$, showing triangles 1 and 2 of spins rotating counter-clockwise for $D<0$, and (b) the HC lattice with sites 1 through 6 indicated.  The solid arrows give GS(0)
while the lighter-colored dash-dot arrows for spins 2, 4, and 6 give GS($\pi/3$).
}
\label{Fig1}
\end{figure}

This paper is divided into 6 sections.  Section II treats the properties of GS($\phi )$ and describes the minimization of the zero-point energy.  
Section III provides results for the DOS in an applied field and identifies the five phases of GS(0).   
The BC and CN of both GS(0) and GS($\pi/3$) are described in Section IV, which also shows that the CN is non-integer
when $\phi $ is not a multiple of $\pi /3$.  INS results are presented in Section V.
A discussion and conclusion are provided in Section VI.   Using the SW formalism, the Appendix describes how to avoid the major source of 
numerical error when evaluating the BC and CN.

\section{Ground State}

The  Hamiltonian for this problem  is given by 
\begin{eqnarray}
{\cal H}&=&-J\sum_{\langle i,j\rangle } \,\vS_i \cdot \vS_j -D\sum_{i,j}(\vS_i\times \vS_j)\cdot \vz \nonumber \\
&-& K\sum_i {S_{iz}}^2 - g\mu_{\rm B} H \sum_i S_{iz},
\end{eqnarray}
where the FM exchange interaction $J>0$ is summed over nearest-neighbor sites and the DM interaction $-D(\vS_i\times \vS_j)\cdot \vz $ is taken over
next-nearest-neighbor sites and 
oriented from sites 1 to 3, sites 3 to 5, and sites 5 to 1 on triangle 1 and from sites 2 to 4, sites 4 to 6, and sites 6 to 2 on 
triangle 2.  Both the easy-axis anisotropy $K$ and the magnetic field $H$ lie along $\vz $. 

For any field, the spins can be parameterized as
\begin{equation}
\vS_1(\theta , \phi ) = S \bigl(\sin \theta ,0 ,\cos \theta \bigr),
\end{equation}
\begin{equation}
\vS_2(\theta , \phi ) = S \bigl(\sin \theta \cos \phi ,\sin \theta \sin \phi ,\cos \theta \bigr),
\end{equation}
\begin{equation}
\vS_3(\theta, \phi ) = S \bigl(\sin \theta \cos (2\pi/3) ,\sin \theta \sin (2\pi/3)  ,\cos \theta \bigr),
\end{equation}
\begin{eqnarray}
\vS_4(\theta,  \phi ) &=& S \bigl(\sin \theta \cos (2\pi/3 + \phi ) ,\sin \theta \sin (2\pi/3 +\phi) ,\nonumber \\
&&\cos \theta \bigr),
\end{eqnarray}
\begin{equation}
\vS_5(\theta , \phi ) = S \bigl(\sin \theta \cos (4\pi/3) ,\sin \theta \sin (4\pi/3)  ,\cos \theta \bigr),
\end{equation}
\begin{eqnarray}
\vS_6(\theta , \phi ) &=& S \bigl(\sin \theta \cos (4\pi/3 +\phi ) ,\sin \theta \sin (4\pi/3 + \phi )  ,\nonumber \\
&&\cos \theta \bigr).
\end{eqnarray}
Due to the twisting angle $\phi $, the spins $\{1,3,5\}$ on triangle 1 are out of phase with spins $\{2,4,6\}$ on triangle 2 as seen in Fig.\,1(a).  Regardless of $\phi $,
spins on each triangle $\{1,3,5\}$ or $\{2,4,6\}$ rotate counter-clockwise by $2\pi /3$ or 120$^o$.  
The spin state given above can be roughly described as a cone with spins tilted 
towards the $z$ axis.  The $xy$ spin configurations with $\phi =0$ and $\pi/3$ are plotted in Fig.\,1(b).  

With fixed $\theta $, the spin configuration above minimizes the energy $E=\langle {\cal H}\rangle $ 
evaluated for classical spins with one independent angle per spin.  
Of course, the energy is degenerate with respect to an overall rotation of the spin state about the $z$ axis.  
Since this rotation does not cost any energy, it is associated with a Goldstone mode at the ordering wavevector $\vQ = (0,2\sqrt{3}/9)(2\pi/a)$ in the INS spectrum 
$S(\vk ,\omega )$ presented in Section V.

To order $S^2$, the energy $E_0$ of the above spin configuration is given by
\begin{eqnarray}
\frac{E_0}{N}&&=-\frac{3}{2}JS^2\cos^2\theta  - KS^2 \cos^2 \theta \nonumber \\
&& +\frac{3\sqrt{3}}{2}DS^2\sin^2\theta  -g\mu_{\rm B}H S\cos\theta ,
\label{E0}
\end{eqnarray}
where $N$ is the number of sites.  Notice that $E_0/N$ is independent of the twist angle $\phi $ at any field.  This can be understood by considering the 
exchange environment of any individual spin $\vS_i$.  Regardless of the orientation of $\vS_i$, the three neighboring spins form one of the two triangles $\{1,3,5\}$
or $\{2,4,6\}$ in Fig.\,1(a) with no net moment in the $xy$ plane.  Hence, their total exchange interaction with $\vS_i$ cancels out within the plane and is independent of $\phi $.
If the field energy $g\mu_{\rm B}H$ is considered to be of order $JS$, then all terms within the classical GS energy $E_0$ are of order $S^2$.

\begin{figure}
\begin{center}
\includegraphics[width=8.5cm]{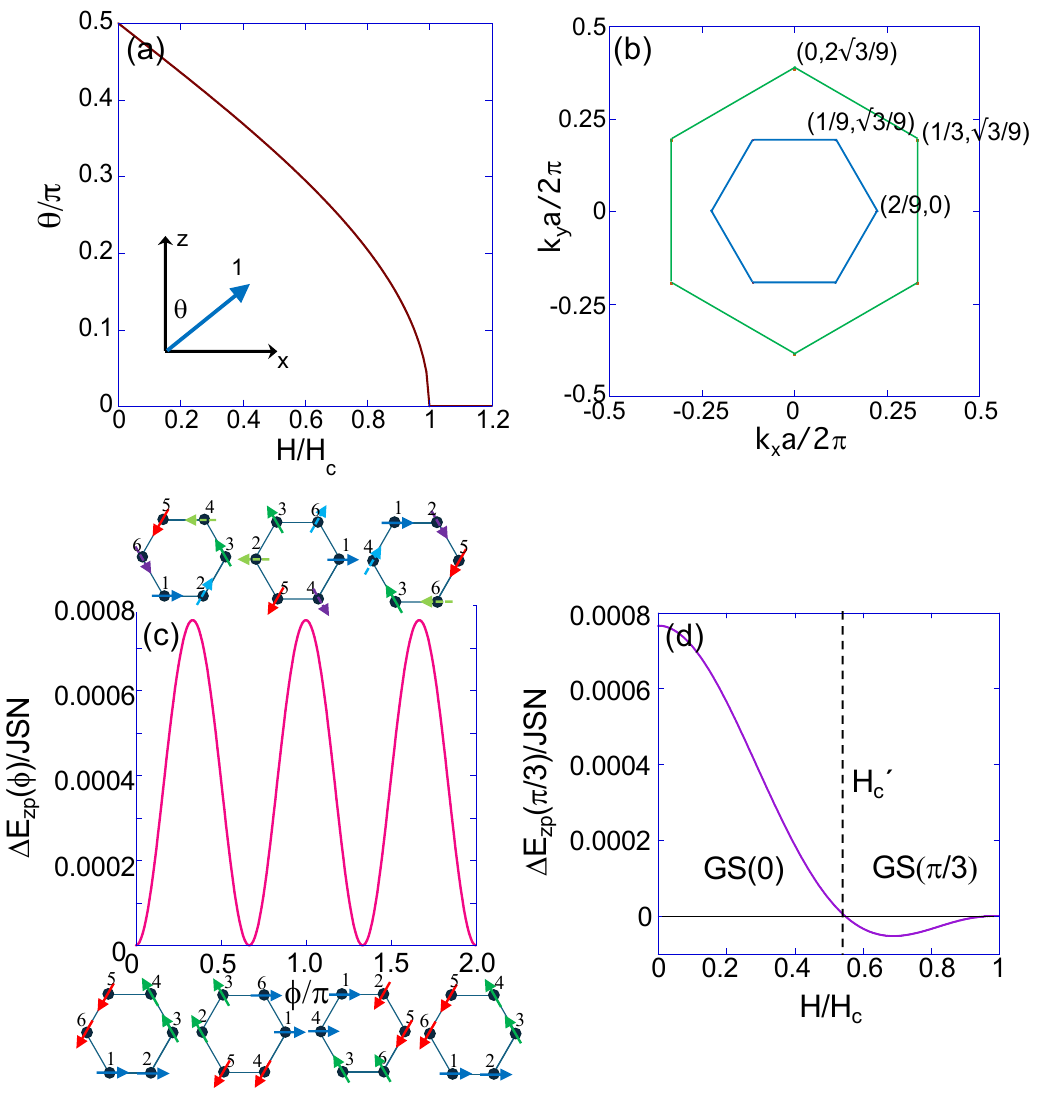}
\end{center}
\caption{(a) The tilting angle $\theta $ versus normalized magnetic field $H/H_c$.  (b) The first BZ for the six subband model is plotted in blue; the first BZ of the two subband model is plotted in green. 
(c) The magnon zero-point energy versus $\phi $ in zero field.   $\Delta E_{\rm zp}(\phi ) = E_{\rm zp}(\phi )-E_{\rm zp}(0)$
where $E_ {\rm zp}(0)$ is the zero-point energy for $\phi = 0$.  On the bottom of Fig.\,2(b) are the GSs for $\phi =0$, $2\pi /3$, $4\pi/3$, and $6\pi/3$.  On the top
are the GSs for $\pi/3$, $\pi $, and $5\pi/3$.  (d)  The difference $\Delta E_{\rm zp}(\pi /3) =E_{\rm zp}(\pi/3)-E_{\rm zp}(0)$ versus $H/H_c$ showing that GS($\pi/3$) has lower energy than GS(0) 
above $H_c'=0.55 H_c$.}
\label{Fig2}
\end{figure}

\begin{figure*}
\begin{center}
\makebox[\textwidth]{\includegraphics[width=.75\paperwidth]{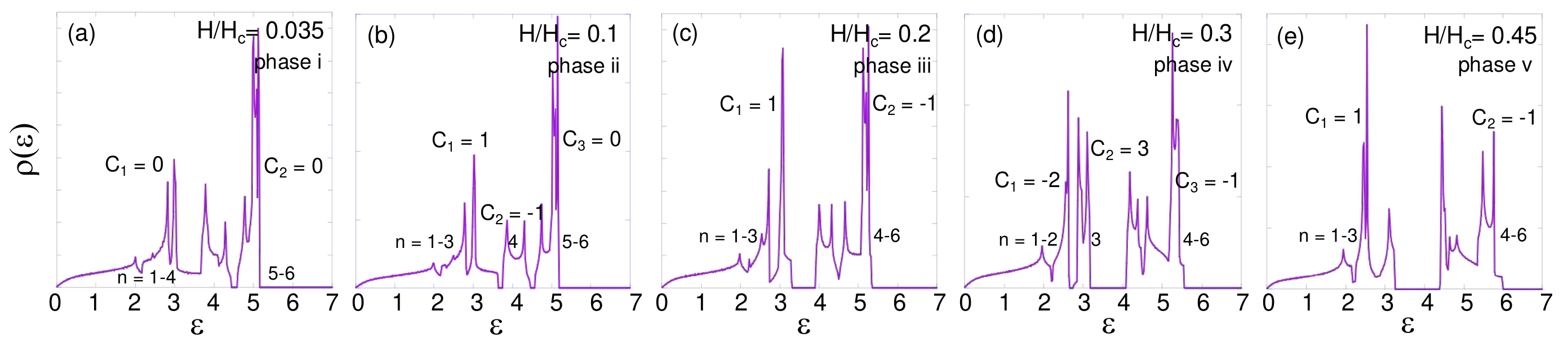}}
\end{center}
\caption{The magnon DOS $\rho (\epsilon )$ versus $\epsilon =\hbar \omega /JS$ for GS(0) with $H/H_c$ = (a) 0.035 ($\theta /\pi = 0.490$), (b) 0.1 ($\theta /\pi = 0.469$),
(c) 0.2 ($\theta /\pi = 0.436$), (d) 0.3 ($\theta /\pi = 0.404$), and (e) 0.45 ($\theta /\pi = 0.351$) using $K/J=0.4$ and $D/J=-0.8$.}
\label{Fig3}
\end{figure*}

Stability of the coplanar states in Fig.\,1(b) requires that $D < -D_c <0$ (positive $0<D_c  < D$ would cause the spins on
each triangle $\{1,3,5\}$ or $\{2,4,6\}$ to rotate clockwise rather than counter-clockwise) where 
\begin{equation}
D_c  = \frac{\sqrt{3}}{3}\biggl\{ J +\frac{2}{3}K \biggr\} .
\label{dc}
\end{equation}
The field dependence of $\theta $ is then given by
\begin{equation}
\cos \theta =\frac{g\mu_{\rm B}H}{2GS},
\label{ca}
\end{equation}
with
\begin{equation}
G=-\frac{3}{2}J-K+\frac{3\sqrt{3}}{2}\vert D\vert 
= \frac{3\sqrt{3}}{2} \Bigr\{ \vert D\vert - D_c  \Bigr\} > 0.
\end{equation}  
The critical field corresponding to $\theta =0$ is 
\begin{equation}
H_c=\frac{2GS}{g\mu_{\rm B}}
=\frac{3\sqrt{3}S}{g\mu_{\rm B}} \Bigr\{ \vert D\vert - D_c\Bigr\}.
\end{equation}    
We plot $\theta /\pi $ versus $H/H_c$ in Figure 2(a).  Below $H_c$, the magnetization per site is
\begin{equation}
m=g\mu_{\rm B}S\cos \theta =\frac{(g\mu_{\rm B})^2H}{3\sqrt{3}\bigl(\vert D\vert -D_c \bigr)}
=g\mu_{\rm B}S\, \frac{H}{H_c},
\end{equation}
which is linear in field.

As $\vert D\vert \rightarrow D_c $ with $\vert D\vert >  D_c$, $G\rightarrow 0^+$ and $H_c\rightarrow 0$.  
When $\vert D\vert <  D_c $, $G< 0$ and Eq.\,(\ref{ca}) has no solution for $\theta > 0$.  
Then $\theta =0$, $H_c=0$, and the spins are always aligned along $\vz $.

While the GS energy given by Eq.\,(\ref{E0}) is independent of the twist angle $\phi $, the SW frequencies do depend on $\phi $.  
The SW frequencies can be evaluated to order $S$ in a $1/S$ expansion of the Hamiltonian.  
Since there are $M=6$ sites in the magnetic unit cell below $H_c$, 
we calculate the SW frequencies $\omega_n(\vk )$ by diagonalizing a $2M\times 2M = 12\times 12$ matrix using the standard techniques \cite{fishmanbook18} briefly described in the Appendix.   
Frequencies are ordered at each wavevector $\vk $ 
from lowest to highest frequency to give subbands $n=1$ through $6$.  

The first BZ corresponding to the magnetic unit cell of the six-subband model 
is the hexagon with area $(2\sqrt{3}/27)(2\pi /a)^2$ plotted in the smaller, blue hexagon of Fig.\,2(b).
For comparison, the first BZ of the two-subband model valid in the FM state with two sites (say sites 1 and 2 in Fig.\,1(b)) in the 
magnetic unit cell is sketched in the larger, green hexagon of Fig.\,2(b).  This hexagon has
area $(2\sqrt{3}/9)(2\pi /a)^2$, which is three times larger than the area of the first BZ hexagon of the six-subband model.  Notice that the two hexagons are rotated by $\pi/2$ 
wrt each other.  We shall use the first BZ of the six-subband model to evaluate the magnon DOS in Section III and the CNs in Section IV.

Henceforth, we shall set $K/J=0.4$ and $D/J=-0.8$ so that $g\mu_{\rm B}H_c = 0.357JS$ and $\vert D\vert  >  D_c  = 0.731J$.  Although this is an unrealistically large value for the DM 
interaction, these parameters are used to construct a model where the DM interaction forces the spins to lie on the $xy$ plane in zero field.  The field then gradually lifts the spins out of the $xy$ plane
towards the $z$ axis.

At temperature $T$, the magnon free energy ${\cal F}$ is given by
the well-known SW result \cite{Kubo52}
\begin{eqnarray}
{\cal F}&=&E-T{\cal S} = \frac{\hbar }{2}\, {\sum_{n, \vk }}'\omega_n(\vk ) \nonumber \\
&+& T\,{\sum_{n,\vk }}'  \log \bigl\{1 - \exp (-\hbar \omega_n(\vk )/T)\bigr\},
\label{FE} 
\end{eqnarray}
where ${\cal S}$ is the entropy.
Summations involve integrals over the first BZ (indicated by the prime) and a sum over the magnon band index $n$.  
The first term in Eq.\,(\ref{FE}) is the zero-point energy $E_{\rm zp}$ due to
quantum SW fluctuations out of the vacuum.  
The normalized zero-point energy per site can be written as 
\begin{equation}
\frac{E_{\rm zp}}{NJS}=\frac{\hbar }{2JSMA_{\rm BZ}}\, \sum_{n=1}^M \, \int_{\rm BZ}d^2k \,\omega_n(\vk ),
\end{equation}
where $A_{\rm BZ}=(2\sqrt{3}/9)(2\pi /a)^2$ is the area of the first BZ and $M=6$ is the number of magnetic sites.
This term dominates the dependence of the normalized free energy per site ${\cal F}/NJS$ on $\phi $ at low temperatures $T/J \ll 1$.   

At zero temperature and zero field, $E_{\rm zp}/NJS$ is evaluated as a function of the twist
angle $\phi $ in Fig.\,2(c).  As shown, $E_{\rm zp}(\phi )/NJS$ has minima at $\phi =\pi l/3 $ for any even integer $l$.  
Spin configurations with $l=0$, 2, and 4 are sketched on the bottom of Fig.\,2(b).
Although $l=0$ ($\phi =0$) is used to construct the GS in Fig.\,1(b), all GSs with even $l$ are equivalent.
With FM nearest-neighbor interactions and easy-axis anisotropy,
GS(0) can also be stabilized by pseudodipolar interactions \cite{Wang21} rather than by the zero-point energy in the presence of DM interactions.

When the field increases, the sign of $(E_{\rm zp}(\pi/3)-E_{\rm zp}(0))/NJS$ flips from positive to negative, as seen in Fig.\,2(d).  Below the crossing field 
$H_c' \approx 0.55 H_c$, the GS has $l=0$ ($\phi =0$);  above $H_c'$,
the GS has $l=1$ ($\phi =\pi/3$).  Spin configurations with $l=1$, 3, and 5 are sketched on the top of Fig.\,2(b).
GS($\pi/3$) with $l=1$ and $\phi =\pi/3$ is plotted by the lighter-colored dashed-arrows for spins 2, 4, and 6 in Fig.\,1(b).  
However, all GSs with odd $l$ are equivalent.  

Notice that the zero-point energy $E_{\rm zp}\sim S$
is two orders of $1/\sqrt{S}$ smaller than the classical GS energy $E_0 \sim S^2$.
While GS(0) has parallel spins $\{1,2\}$, $\{3,4\}$, and $\{5,6\}$, GS($\pi/3$) has anti-parallel spins $\{1,4\}$, $\{2,5\}$, and $\{3,6\}$.
The selection of a state with $\phi =\pi l/3$ for integer $l$ and parallel spins supports the idea that 
collinear spins maximize the correlation of fluctuations \cite{McClarty14}.

\section{Magnon Density of States}

\begin{figure}
\begin{center}
\includegraphics[width=8.5cm]{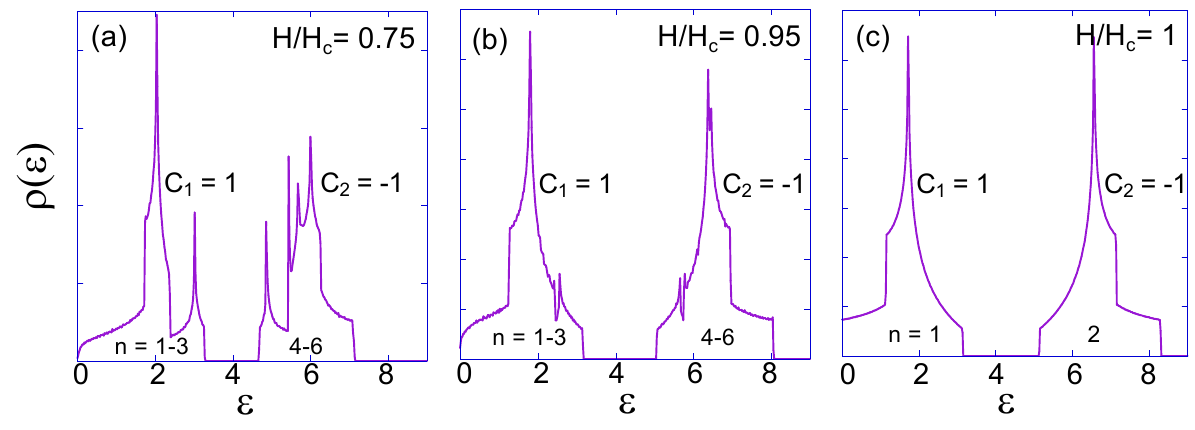}
\end{center}
\caption{The magnon DOS $\rho (\epsilon )$ versus $\epsilon =\hbar \omega /JS$ for GS($\pi/3$) with $H/H_c$ = (a) 0.75 ($\theta /\pi = 0.230$), (b) 0.95 ($\theta /\pi = 0.101$), and
(c) 0 ($\theta /\pi = 0$).  Other parameters same as in Fig.\,3.}
\label{Fig4}
\end{figure}

To obtain the magnon DOS, we evaluate the frequencies at each $\vk $ point on a grid of the first BZ, labelling them as subbands $n=1$ through 6 in order of increasing 
energies $E=\hbar \omega_n(\vk )$.  We then count the number of $\vk $ points that have energies within an interval between 
$E_i$ and $E_j=E_i+\Delta E$ where $E_i$ and $E_j$ extend from 0 to $9J$, which allows for all possible
eigenvalues at all fields.  $\Delta E$ is chosen sufficiently small (of order $0.01J$) to give a smooth DOS.  Finally, we normalize 
the DOS $\rho (\epsilon )$ so that
\begin{equation}
\int_0^{\infty } d\epsilon\, \rho(\epsilon )=1,
\end{equation}
where $\epsilon =E/JS$.

For future discussion, {\it bands} shall refer to the set of $N_b$ magnonic states that are isolated by energy gaps from one another.   
A {\it subband} shall refer to an excited state produced by one of the $M$
eigenstates of the Hamiltonian.  A single band can contain up to $M$ subbands.  While the number $N_b$ of bands can vary, the number of subbands is 
fixed by the number $M$ of magnetic sites in the unit cell.
A subband only ever belongs to a single band.  In our model, $M=6$ and (as we shall see) $N_b=2$ or 3.

Results for the DOS of GS(0) are given in Figure 3.  This figure allows us to identify 5 distinct phases that appear due to changes in the DOS.  
In phase $i$ (including zero field), there are $N_b=2$ bands.  Band 1 consists of subbands 
1-4 while band 2 consists of subbands 5-6.  The small gap between these bands can be seen in Fig.\,3(a).  This phase is centered around $H/H_c = 0.035$ but there is little difference
between the DOS from zero field to that pictured in Fig.\,3(a).
In phase $ii$, an additional gap appears between subbands 1-3 and subband 4, producing a total of $N_b=3$ bands.   This phase is centered around $H/H_c = 0.1$.
In phase $iii$, the gap between subband 4 and subbands 5-6 closes.  So there is only 1 gap and $N_b=2$ bands remain.  This phase is centered around $H/H_c =0.2$.
In phase $iv$, a new gap appears between subbands 1-2 and subband 3.  With two gaps and $N_b=3$ bands, this phase is centered around $H/H_c = 0.3$.
Finally, the lower gap disappears in phase $v$ leaving only one surviving gap between subbands 1-3 in band 1 and subbands 4-6 in band 2.  This phase is centered around $H/H_c=0.45$ and persists
until $H_c'=0.55H_c$.  

\begin{figure}
\begin{center}
\includegraphics[width=7cm]{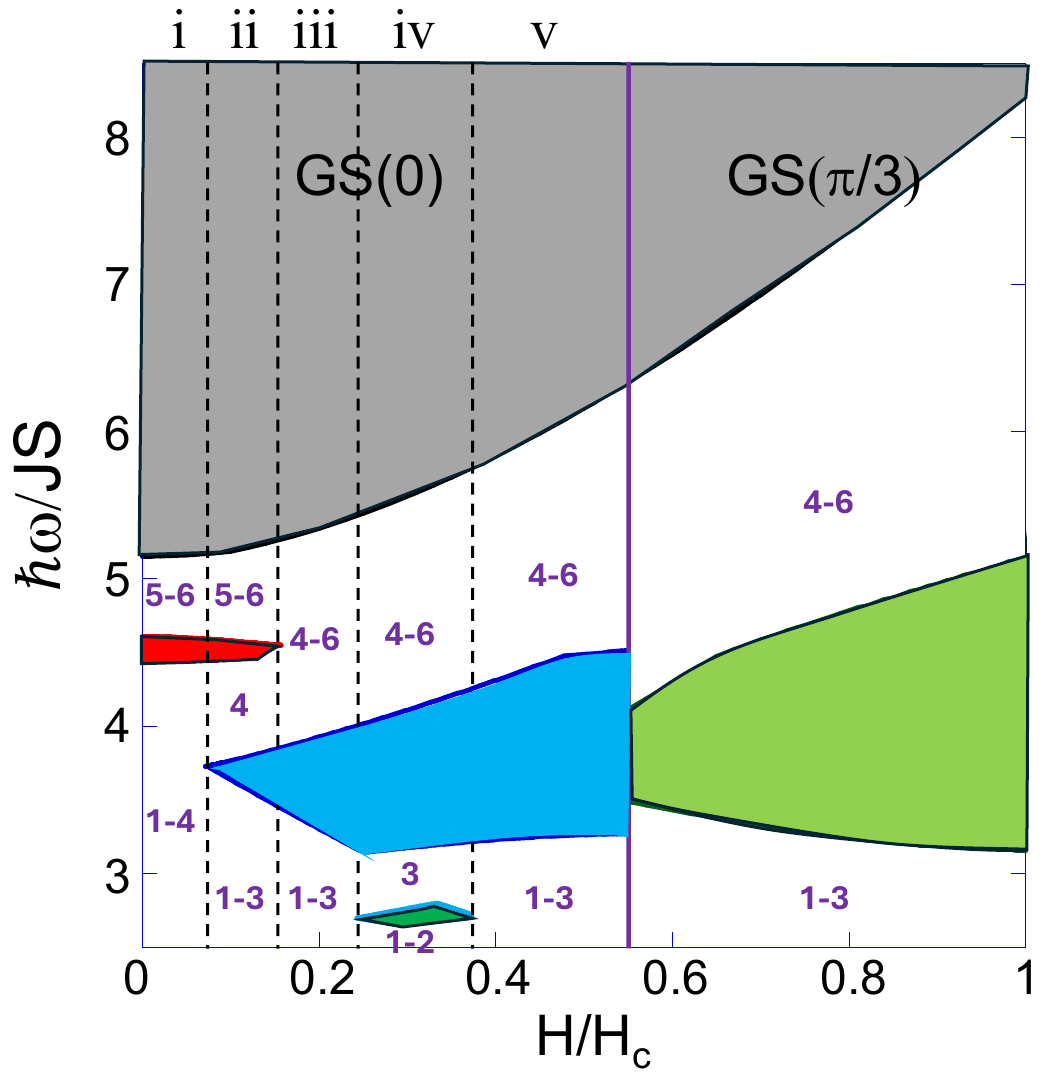}
\end{center}
\caption{The phase diagram for the subband index corresponding to Figs.\,3 and 4.  Numbers 1 through 6 indicate the subbands that contribute to the DOS.  Colored regions
contain no states.  Other parameters the same as for Fig.\,3.}
\label{Fig5}
\end{figure}

\begin{figure}
\begin{center}
\includegraphics[width=8.5cm]{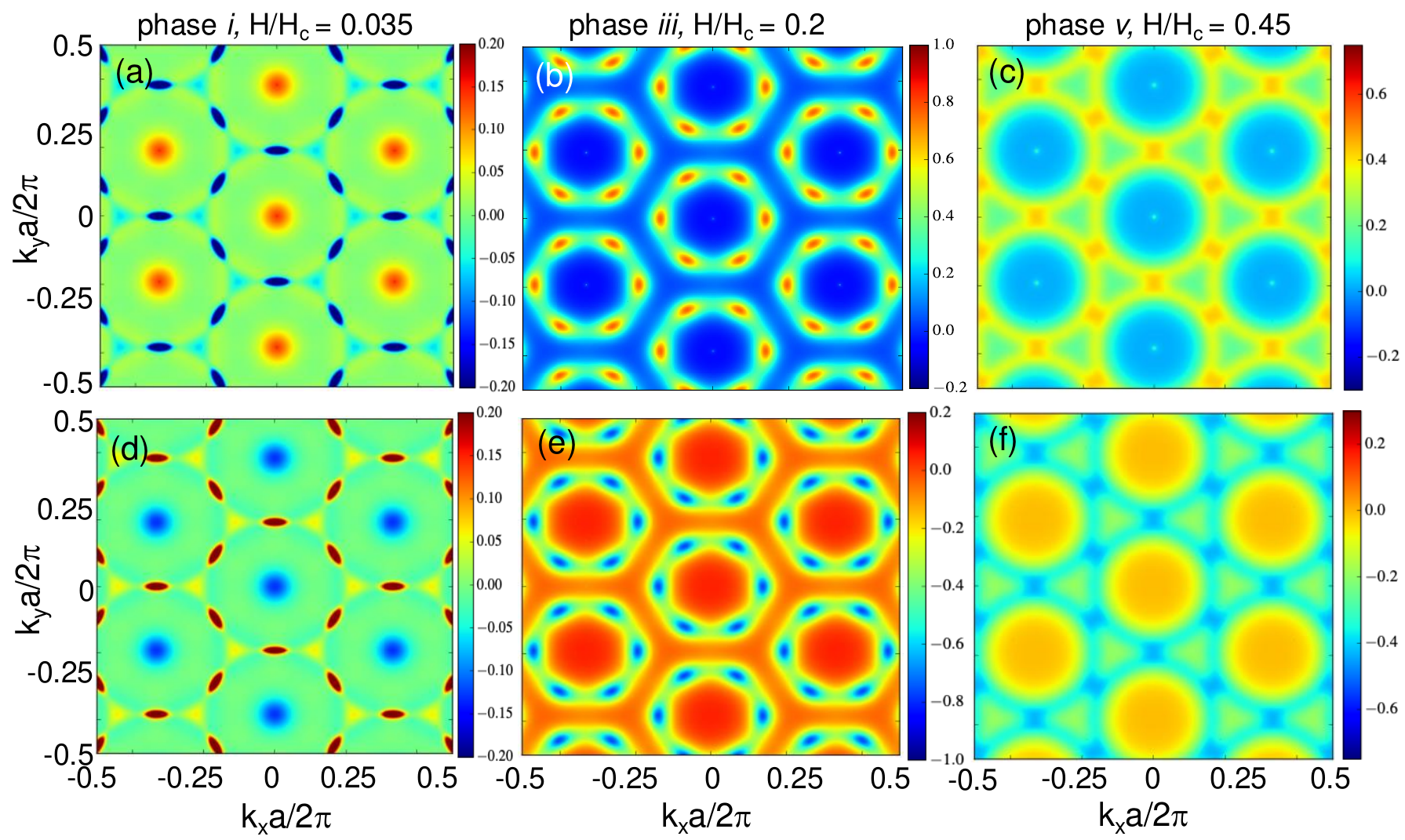}
\end{center}
\caption{The BC $\Omega_{1z}(\vk )$ versus $\vk $ for phases $i$, $iii$, and $v$ with two bands, all within GS(0):  band 1 at the top and band 2 at the bottom.
Fields are $H/H_c=0.035$ in phase $i$ for (a) and (d), $H/H_c=0.2$ in phase $iii$ for (b) and (e), $H/H_c=0.45$ in phase $v$ for (c) and (f).  Other parameters the same as for Fig.\,3.}
\label{Fig6}
\end{figure}

In Figure 4, we plot the DOS for two fields in GS($\pi/3$), $H/H_c=0.75$ and 0.95, in addition to $H/H_c=1$.  For the latter, we use a two-band 
model for the FM, 
which explains why there are only two subbands $n=1$ and 2.  The FM DOS exhibits mirror symmetry about the center of the gap.  This symmetry is absent 
below $H_c$.  Not surprisingly, the DOS for $H/H_c=0.95$ approaches the DOS for the FM aside from two small notches on either side of the gap.  
These notches become larger at the lower field $H/H_c=0.75$.  Also, the top band develops structure that becomes more complex with decreasing field, 
as can be seen for $H/H_c=0.45$ within GS(0) in Fig.\,3(e).

The band phase diagram is plotted in Figure 5.  While colored regions contain no subbands, white regions contain subbands indicated by integers 1 through 6.  Phases 
$i$ through $v$ for GS(0) are indicated on top of the figure.
As seen, the largest and dominant gap lies between subbands 1-3 and subbands 4-6.  At the top of the graph, the grey region indicates that all of the states have been exhausted.
The curve marking the boundary of this region shows no discontinuity at the border between GS(0) and GS($\pi/3$).
On the other hand, the middle gap shrinks by about 50\% upon going from GS(0) to GS($\pi/3$).  
This will have physical consequences for the magnon Hall effect and the edge modes.

\section{Berry Curvatures and Chern Numbers}

As mentioned earlier, evaluating the BC for complex NC states is challenging due to the appearance of the derivative $\partial \vert u_n(\vk )\rangle /\partial \vk $.   
While this derivative can be evaluated analytically for collinear and simple NC states with $M=2$, 
it must usually be evaluated numerically for complex NC states with $M >2$ sublattices.  
Because the eigenvalue equation only determines the eigenstate $\vert u_n(\vk )\rangle $
up to an overall phase, it can then gain a phase factor from one $\vk $ point to the next, which creates a numerical error in the derivative wrt $\vk $.  
In order to avoid this error, we must fix the phase of one component of the eigenvector $\vert u_n(\vk )\rangle $ at every $\vk $ point.  
We demonstrate this technique in the Appendix using the SW formalism.

\begin{figure}
\begin{center}
\includegraphics[width=8.5cm]{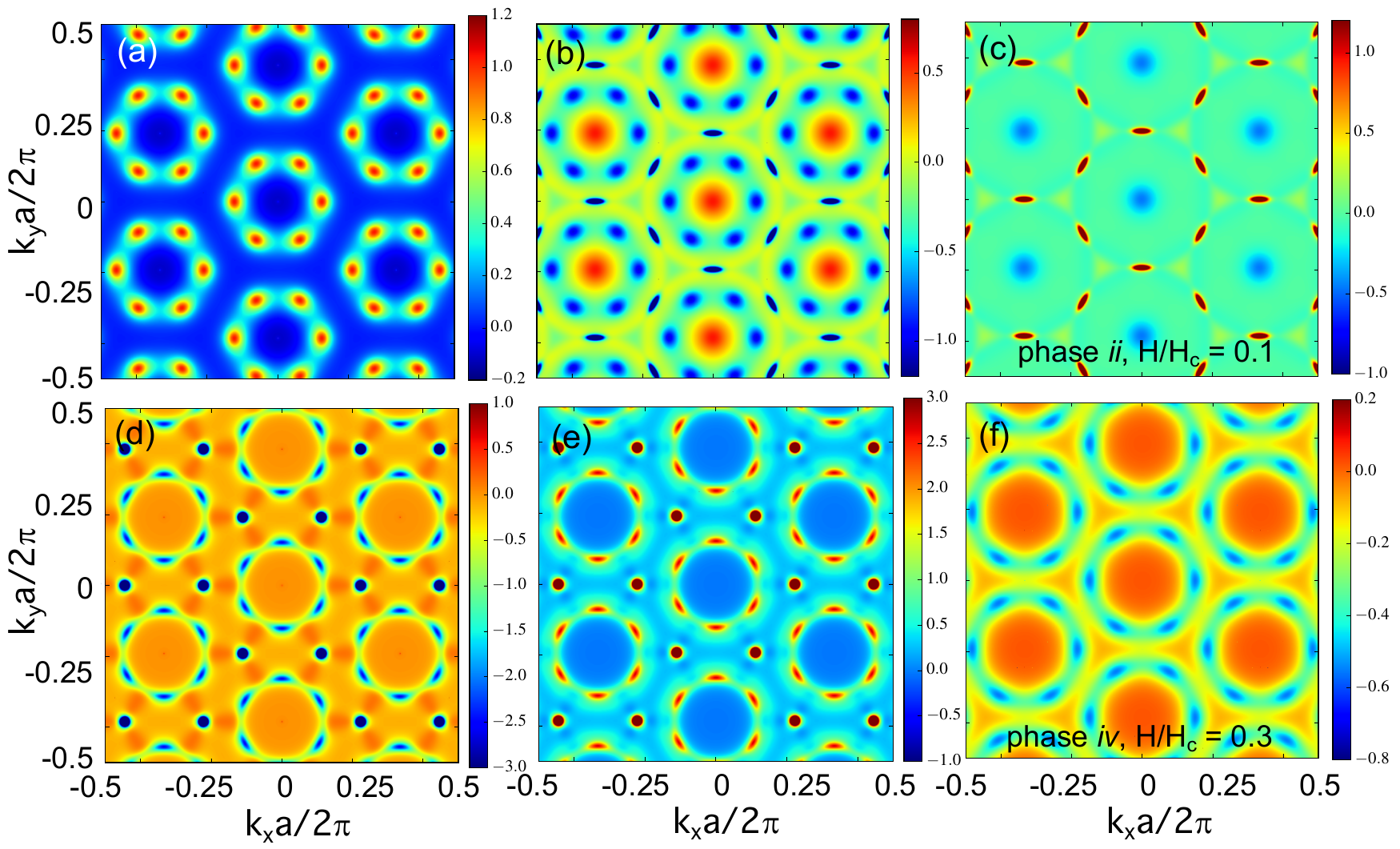}
\end{center}
\caption{The BC $\Omega_{1z}(\vk )$ versus $\vk $ for phases $ii$ and $iv$ with three bands, all within GS(0):  (a), (b) and (c) for bands 1, 2, and 3 in phase $ii$ with $H/H_c=0.1$ and
(d), (e), and (f) for bands 1, 2, and 3 in phase $iv$ with $H/H_c= 0.3$.
Other parameters the same as for Fig.\,3.}
\label{Fig7}
\end{figure}

\begin{figure}
\begin{center}
\includegraphics[width=8.5cm]{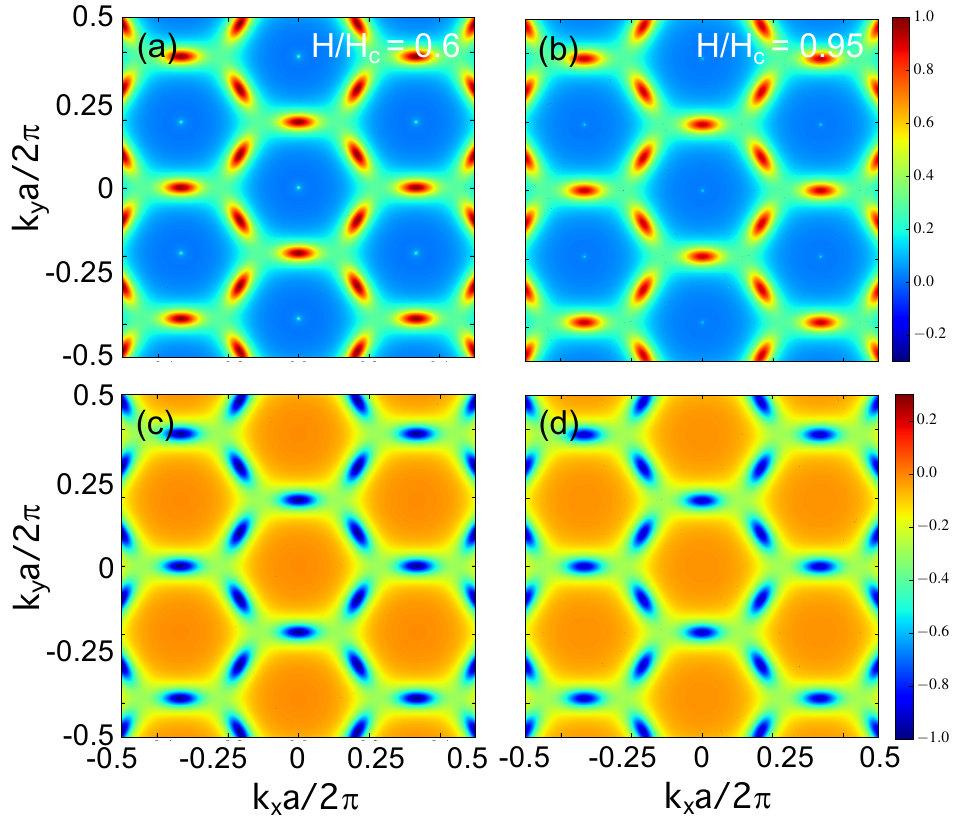}
\end{center}
\caption{The BC $\Omega_{1z}(\vk )$ versus $\vk $ within GS($\pi/3$):  (a) and (c) for $H/H_c=0.6$ and (b) and (d) for $H/H_c=0.95$ with lower (top) and upper (bottom) bands.
Other parameters the same as for Fig.\,3.}
\label{Fig8}
\end{figure}

We evaluate the BC using the same parameters as for the DOS in Figure 3.  When $H=0$, we find that the 
BC vanishes identically for every $\vk $ point.  This also follows from an examination of GS(0), which is symmetric wrt time reversal $\cal{T}$ (a reversal of the spins in the $xy$ plane) 
followed by a $\pi $ rotation about a $z$ axis placed at the center of each bond between sites $\{1,2\}$, $\{3.4\}$, or $\{5,6\}$ (for $\phi =0$)
${\cal C}_2$.  Note that ${\cal{TC}}_2$ symmetry only holds when the twist angle $\phi $ is a multiple of $2\pi /3$ \cite{Vanderbilt18}.  
When $\phi \ne 2\pi l/3$, ${\cal{TC}}_2$ symmetry is broken and the BC is nonzero in the GS.

Compared to the total BC of subbands 1-3 at each $\vk $ point, the total BC of subbands 4-6 has the opposite sign.  
Hence, the total BC of all six subbands is always zero at every $\vk $ point.   This relation holds for all tilt angles $\theta $ within both GS(0) and GS($\pi /3$).
  
\begin{figure}
\begin{center}
\includegraphics[width=7cm]{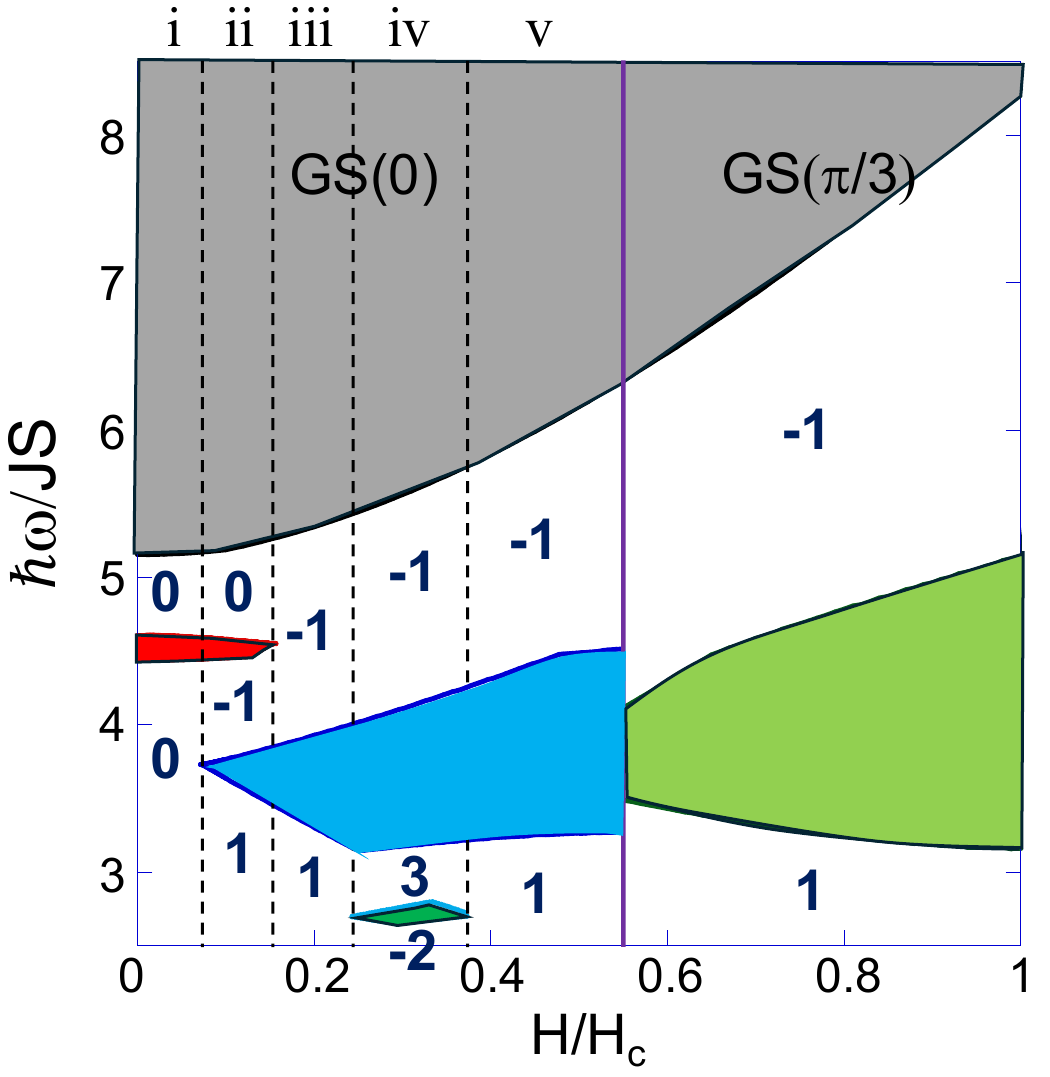}
\end{center}
\caption{The phase diagram for the CNs corresponding to Fig.\,5.  Bands occupy white regions, gaps occupy colored regions.
Parameters the same as for Fig.\,3.}
\label{Fig9}
\end{figure}

Using Eq.\,(\ref{Chern}), we find that the CNs of subband 4 is always -1 and that of subbands 5 and 6 is always 0.  Hence, the total CN of subbands 1-3 is always 1.
This can be verified from Fig.\,3 for the magnon DOS in GS(0).
By using $\sim 10^3$ points within the BZ, the numerically evaluated CNs are precise to within $10^{-5}$.

The BCs for fields with two and three bands are separately plotted for GS(0) in Figures 6 and 7, respectively.  
In Fig.\,6, we plot the BCs for phases $i$, $iii$, and $v$ with two bands.  In Fig.\,7, we plot the BCs for 
phases $ii$ and $iv$ with three bands.

For phase $i$ with $H/H_c=0.035$, the BC of band 1, which consists of subbands 1-4 and yields a CN of 0, is plotted in Fig.\,6(a).   
The BC of band 2, consisting of subbands 5-6 and plotted in Fig.\,6(d), is opposite in sign to band 1 and again yields a CN of 0.  
Within phase $i$, the amplitude of the BC for each band scales like the field.  So the same plots for the BC would be obtained as in Figs.\,6(a) and 
(d) for $H/H_c=0.0175$ if the color scale were changed from $(0.2,-0.2)$ to $(0.1,-0.1)$.  
Consequently, the BC linearly goes to zero as $H\rightarrow 0$ and $\theta \rightarrow \pi /2$.  

\begin{figure}
\begin{center}
\includegraphics[width=8.5cm]{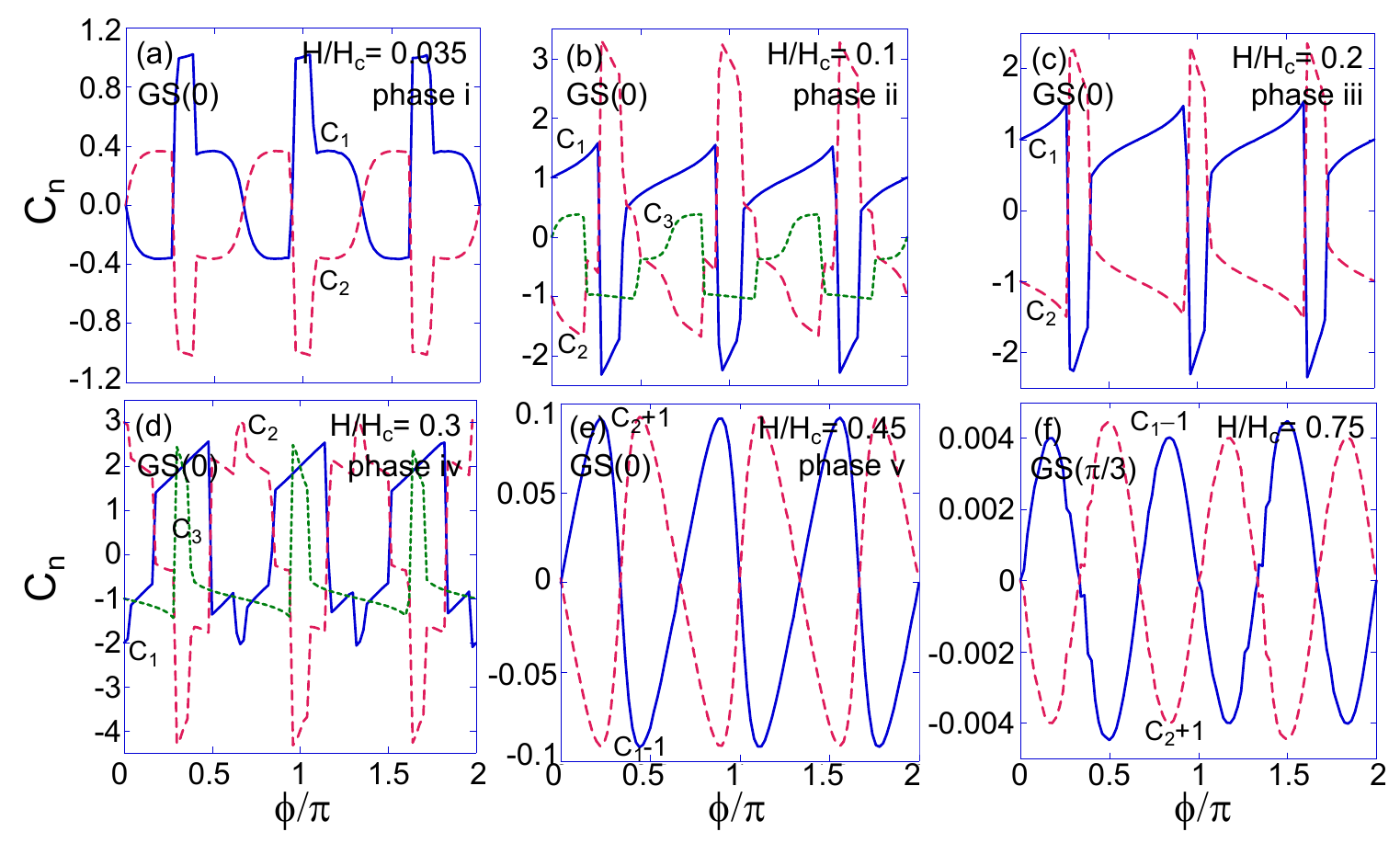}
\end{center}
\caption{The CNs versus twist angle $\phi /\pi $ for phases (a) $i$ with $H/H_c=0.035$, (b) $ii$ with $H/H_c=0.1$, (c) $iii$ with $H/H_c=0.2$, and (d) $iv$ with $H/H_c=0.3$ 
of GS(0).  We then plot the differences $C_1-1 $ and $C_2+1$ for (e) phase $v$ of GS(0) with $H/H_c=0.45$ and (f) GS($\pi/3$) with $H/H_c=0.75$ 
(these results are symmetrized to reduce numerical error).  Other parameters the same as for Fig.\,3.}
\label{Fig10}
\end{figure}

For phase $iii$ with $H/H_c=0.2$, band 1 in Fig.\,6(b) consists of subbands 1-3 with a CN
of 1.  Band 2 consisting of subbands 4-6 and plotted in Fig.\,6(e) has opposite sign and a CN of -1.
Finally, Figs.\,6(c) and (f) for phase $v$ with $H/H_c=0.6$
plot the BCs of band 1 and 2, which consist of subbands 1-3 with a CN of 1 and subbands 4-6 with a CN of -1.

Figure 7 plots the BCs of phases $ii$ and $iv$ with three bands. In phase $ii$, we find that even band 3 with CN of 0 
in Fig.\,7(c) can have a significant BC
as a function of $\vk $.  Consequently, the BCs of bands 1 and 2 with CNs of $+1$ and $-1$, respectively, are not opposite to each other.
In phase $iv$, band 1 with subbands 1-2 has a CN of -2 and band 2 with subband 3 has a CN of +3.  Both these bands have 
large BCs.  By comparison, band 3 consisting of subbands 4-6 has a CN of -1 and a relatively small BC.  Of course, the CNs of these bands
would switch signs if the field were reversed.

As expected from the different band structures, the BCs are remarkably different in each of the five phases of GS(0).
The richness of the DOS and BC should also appear in the variation of the edge states of the HC lattice as the magnetic field increases, 
particularly in phase $iv$.

Figure 8 plots the BCs of GS($\pi/3$) with $H/H_c=0.6$ and 0.95.  These BCs hardly change across GS($\pi/3$) with increasing field, 
except that the small dips in intensity at the center of each hexagon disappear in the FM state.  Also notice that the intensity at the center of 
each oval softens as the field increases.

\begin{figure*}
\begin{center}
\makebox[\textwidth]{\includegraphics[width=.75\paperwidth]{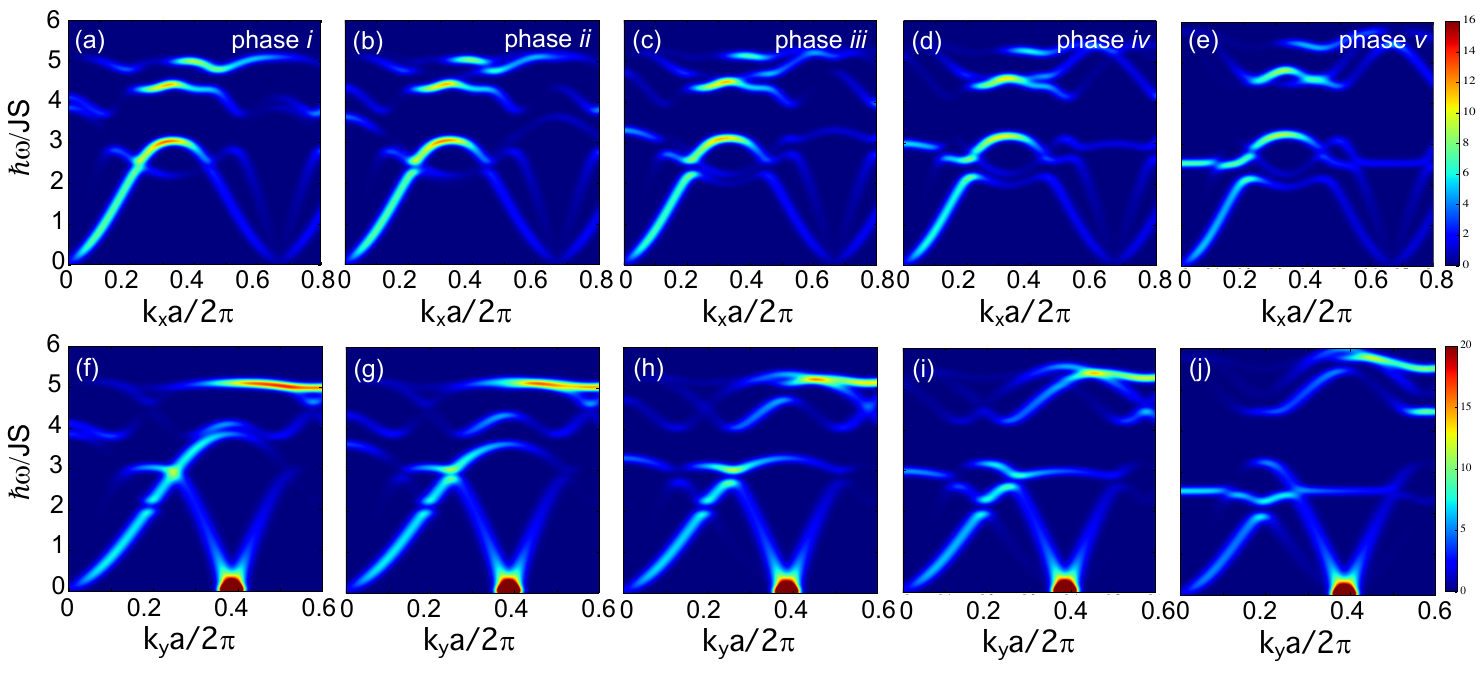}}
\end{center}
\caption{The inelastic spectrum $S(\vk ,\omega )$ corresponding to all five phases of GS(0) with $k_y=0$ (top) and $k_x=0$ (bottom) and 
$H/H_c=$ (a) and (f) $0.035$ in phase $i$, 
(b) and (g) $0.1$ in phase $ii$, (c) and (h) $0.2$ in phase $iii$, (d) and (i) $0.3$ in phase $iv$, and (e) and (j) $0.45$ in phase $v$.  
Other parameters the same as for Fig.\,3.}
\label{Fig11}
\end{figure*}

We plot the phase diagram for the CN rather than the subband index in Figure 9.  Obviously, the only way for a two-band system
with CNs of 0 to evolve into a two-band system with CNs of $\pm 1$ while conserving the total CN across phase boundaries is to pass through an intermediate three-band system.
Notice how the total CN remains conserved across each phase boundary in Fig.\,9:  0 transforms into 1 and -1, -1 transforms into -1 and 0, 1 transforms into -2 and 3, and so on.
Within GS(0), our model evolves from two bands ($C_1=C_2=0$) to three bands ($C_1=1$, $C_2=-1$, $C_3=0$) to two bands ($C_1=1$, $C_2=-1$) to three bands
($C_1=-2$, $C_2=3$, $C_3=-1$) back to two bands ($C_1=1$, $C_2=-1$).  It retains two bands ($C_1=1$, $C_2=-1$) in GS($\pi/3$).

These results assume that the twist angle $\phi $ is a multiple of $\pi /3$.  If $\phi \ne \pi l/3$, then the CNs in 
zero field are nonzero and non-integer and
remain non-integer as the system evolves with field.  This is demonstrated in Figure 10, where we plot the CNs of the magnon
bands as a function of $\phi /\pi $ for each of the five phases in GS(0) and for GS($\pi/3$).  The CN jumps when the twist angle 
changes the DOS by opening or closing a band gap.  For all twist angles, the total CN is zero.  Except when $\phi = \pi l/3$, 
the CNs are non-integer.   For low fields as in Fig.\,10(a) for phase $i$ of GS(0), the CNs return to the same $\phi =0$ values when $\phi =2\pi /3$ or $4\pi/3$, 
as expected from the discussion of the zero-point energy.  For phase $v$ of GS(0) as in Fig.\,10(e) and for GS($\pi/3$) as in Fig.\,10(f), $C_1-1$ and $C_2+1$ 
have periods of $\pi /3$ with amplitudes that decrease as $H\rightarrow H_c$.  So the twist angle $\phi $ becomes irrelevant as $H\rightarrow H_c$ or as 
GS($\pi/3$) approaches a FM.  
We emphasize once again that twist angles $\phi \ne \pi l/3$ are nonphysical below $H_c$.

For a HC lattice with anisotropy $K$ set to zero and AF coupling $J< 0$, Chen {\it et al.} \cite{Chen23} recently studied the evolution of an 
AF with spins along the $x$ axis to a FM 
with spins along $\vz $ in a magnetic field.  While they did not 
find any intermediate phases, their model only contains two magnetic sublattices and two magnon subbands.

\section{Inelastic Neutron-Scattering Results}

We now present the INS spectra $S(\vk ,\omega )$ for all five phases of GS(0) and for GS($\pi/3$) using the SW approximation.  
The inelastic spectra is evaluated in terms of the spin-spin correlation function
\begin{eqnarray}
\label{ssc}
\Sab &=& \frac{1}{2\pi N } \int dt \, e^{-i\omega t} \sum_{i,j} e^{-i\vk \cdot (\vR_i -\vR_j )}\nonumber \\
&&\langle S_{i\alpha }(0)S_{j\beta }(t)\rangle ,
\end{eqnarray}
where $S_{i\alpha }(t)=\exp(i{\cal H}t) S_{i\alpha }\exp(-i{\cal H}t)$.
Including the kinematical constraint that neutrons only couple to spin fluctuations perpendicular to their change in momentum, we 
obtain \cite{fishmanbook18}
\begin{equation}
\label{sqw}
\Svk =\sum_{\alpha , \beta }\Biggl\{ \delta_{\alpha \beta } - \frac{k_{\alpha }k_{\beta }}{k^2} \Biggr\} S_{\alpha \beta }(\vk ,\omega ).
\end{equation}
Because we employ a $1/S$ expansion, each SW mode is associated with a delta function $\delta (\omega -\omega_n(\vk ))$ in $\Svk $.
In the plots below, we broaden those delta functions using reasonable values for the SW lifetimes.
 
As previously, 
we take $K/J=0.4$ and $\vert D\vert /J=0.8$.  Because there are six sites in the magnetic unit cell, 
there are six possible magnon modes per $\vk $ point in the first BZ.   It is important to keep in mind that the number of magnetic modes is 
related to the number of magnetic subbands (six) and not the number of connected bands (two or three), which determine the structure of the 
DOS and of the BC and CN.

In Figure 11 for GS(0), we take $k_y=0$ for the top and $k_x=0$ for the bottom.  
Although only four active SW modes survive when $k_y=0$ and $k_x\ne 0$ at zero field (as in Fig.\,11(a)), 
all six modes appear when $k_x=0$ and $k_y \ne 0$ (as in Fig.\,10(f)). 
Along the $k_x$ axis, the number of active SW modes rises to six at intermediate values of $\theta $ (as in Figs.\,11(b-e)).


The zero-frequency Goldstone mode at $k_ya/2\pi =2\sqrt{3}/9 \approx 0.385$ in the bottom five figures is a signature of the 
120$^o$ order within the $xy$ plane at the ordering wavector $\vQ =(0,2\sqrt{3}/9)(2\pi/a)$.  
Although its intensity diverges as $\omega \rightarrow 0$, this mode gradually weakens as the field increases and the $xy$ order disappears \cite{fishmanbook18}.
The local stability of GS(0) is verified by the reality of the SW frequency $\omega_1(\vk )$ at $\vk = \vQ $.

A strong mode at low fields is seen at $k_xa/2\pi \approx 0.33$ and $\hbar \omega \approx 4.5 JS$.  This mode remains active in 
phases $ii$, $iii$, and $iv$ but mostly disappears in phase $v$, where a new mode appears at $\vk =0$ and $\hbar \omega \approx 2JS$.  
A similar mode appears for $k_y a/2\pi \approx 0.5$ at low fields and $\hbar \omega \approx 5JS$.  
It too rises in frequency and drops in intensity in phase $v$ to be replaced by the 
horizontal mode at $\vk =0$ and $\hbar \omega \approx 2JS$.   

In Figure 12, we plot the INS intensity for GS($\pi/3$) at fields $H/H_c=0.75$, 0.95, and 1 (the FM state).  
As above, the local stability of GS($\pi/3$) is verified by the reality of the Goldstone mode frequency $\omega_{n=1}(\vQ )=0$ below $H_c$.

Because the FM can be described by a two-site model, it only supports two magnon modes with lower and upper
frequencies given analytically by \cite{Fishman23}
\begin{equation}
\hbar \omega_1(\vk)=3JS ( 1  - \eta_{\vk } +\kappa ) +2\mu_{\rm B}H
\label{om1},
\end{equation}
 \begin{equation}
\hbar \omega_2(\vk)=3JS ( 1  + \eta_{\vk } +\kappa ) +2\mu_{\rm B}H,
\label{om2}
\end{equation}
where $\eta_{\vk }=\sqrt{\vert \Gamma_{\vk }\vert^2 +(d\, \Theta_{\vk })^2}$,
$d=-D/3J$, $\kappa = 2K/3J$,
\begin{equation}
\Theta_{\vk} = 4\cos(3k_xa/2) \sin(\sqrt{3} k_ya/2)-2\sin(\sqrt{3}k_ya),
\end{equation}
and
\begin{equation}
\Gamma_{\vk} =\frac{1}{3}\Bigl\{ e^{ik_xa}+2e^{-ik_xa/2} \cos(\sqrt{3}k_ya/2) \Bigr\},
\end{equation}
defined so that $\Gamma_{\vk =0}=1$.
For $H=H_c$, these two modes are seen in Figs.\,12(c) and (f).

\begin{figure}
\begin{center}
\includegraphics[width=8.5cm]{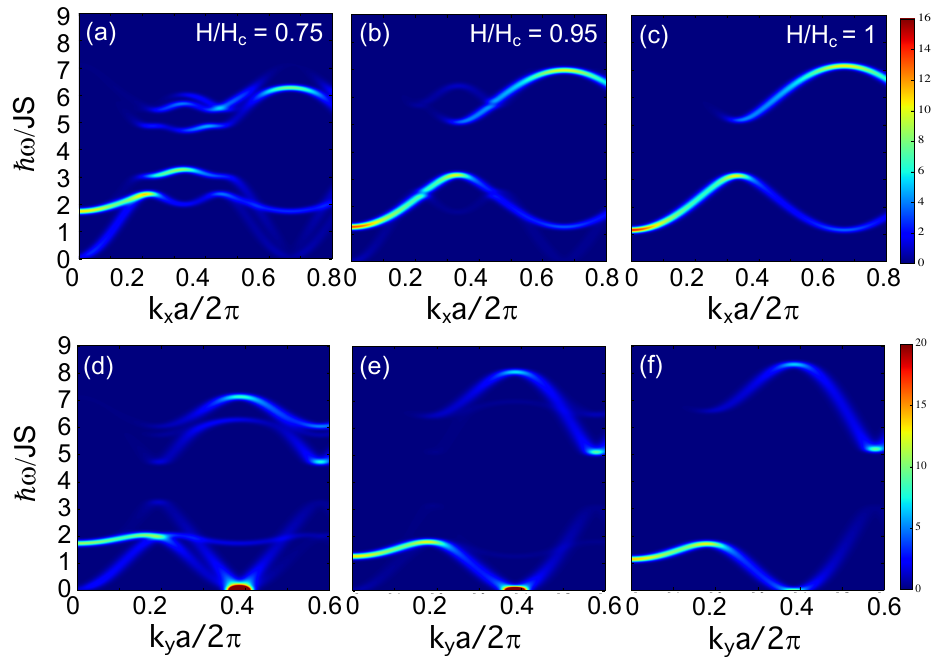}
\end{center}
\caption{The inelastic spectrum $S(\vk, \omega )$ for GS($\pi/3$) with $k_y=0$ (top) and $k_x=0$ (bottom) and $H/H_c=$ (a) and (d) $0.075$, 
(b) and (e) $0.95$, and (c) and (f) $1$.   Other parameters the same as for Fig.\,3.}
\label{Fig12}
\end{figure}

The nearly horizontal $2JS$ modes discussed above at small $\vk $ for GS(0) are seen quite clearly in 
GS($\pi/3$) when $H/H_c=0.75$ and 0.95 in Fig.\,12 and drop in frequency with increasing field.
In the FM at $H_c$, the $\vk =0$ value of this doubly-degenerate mode is given by $\hbar \omega (\vk =0)  = 2KS + 2\mu_{\rm B}H$.
As $H\rightarrow H_c$, the nearly horizontal modes in GS($\pi/3$) approach the $\vk=0$ limit $\hbar \omega (\vk =0) =1.17JS$, 
using the parameters given above.  Of course, the INS intensity no longer diverges at the ordering wavector $\vQ $ at $H=H_c$ in Fig.\,12(f) 
since the spins are aligned along $\vz$.  At fields above $H_c$, the lower mode given by Eq.\,(\ref{om1}) no longer 
vanishes at $\vQ $.

Passing from phases $i$ through $iv$ in GS(0), the largest change in the SW spectrum is the development of a gap at about 
$k_xa/2\pi \approx 0.2$ or $k_ya/2\pi \approx 0.2$ and $\hbar \omega \approx  2.5JS$.   
But generally, the SW spectrum remains relatively unchanged until phase $v$.  
The changes in the spectrum through GS($\pi/3$) are much more dramatic as the SW spectrum simplifies with many of the modes
loosing intensity with increasing field.  It is interesting that the INS spectrum in Fig.\,12 changes so dramatically from $H/H_c=0.75$ to 0.95 
while the BC in Fig.\,8 hardly changes from $H/H_c=0.6$ to 0.95.  This indicates that whereas the BC is most sensitive to changes in the 
magnon DOS, the INS is directly sensitive to changes in the magnon number and intensity with field.
The same intensity scale is used for $S(\vk, \omega )$ for the top and bottom subplots in Figs.\,11 and 12 to more easily compare the drop in 
SW intensity of the modes as the field increases.

Upon close examination, it is possible to connect details of the SW spectrum in Figs.\,11 and 12 with the gaps in the DOS in Fig.\,9.  
For example, the red region in Fig.\,9 indicates a gap in the DOS of phases $i$ and $ii$ between $\hbar \omega /JS=4.45$ and 4.6.
A careful inspection of Figs.\,11(a), (b), (f), and (g) will confirm that there is no INS intensity in this region.  
For phase $v$, there is no scattering intensity between about $\hbar \omega /JS = 3.2$ and 4.4, corresponding to the blue
region in Fig.\,9.  However, there are no dramatic changes in the INS intensity upon going from one phase to the next.

\section{Discussion and Conclusion}

This paper has studied the evolution of the BC and CN as a NC spin state evolves with field from coplanar to FM.
Surprisingly, this relatively simple system built on a HC lattice exhibits complex behavior with two distinct GSs obtained by minimizing the zero-point energy
with respect to the twist angle $\phi $ between triangles of neighboring spins and with five distinct topological phases caused by the change in the 
magnon DOS with the opening and closing of energy gaps in the low-field GS.  This complex topological 
behavior occurs despite the continuous evolution of the tilt angle $\theta $ and magnetization $m$ with field and the relatively simple spin state 
that can be roughly described as a cone tilting towards the $z$ axis.

More than one topological phase appears due to the 
distribution of $M>2$ subbands within $N_b <  M$ connected magnetic bands 
that exhibit BCs and CNs.  Transitions between phases occur as the number $N_b$ of bands change.
Although these topological phases may be difficult to distinguish based on INS measurements, they 
should be easy to identify based on measurements of the edge states and the magnon Hall effect.    

Most remarkably, the complex topological phase diagram of this model appears despite the smooth evolution of the tilting angle $\theta $ and magnetization $m$ with field.
The topological properties of the NC phase sensitively depend on the magnonic DOS, which in turn depends on the SW excitation spectrum.  Despite the smooth evolution
of $\theta $ with field, the change in the excitation spectrum with field produces a redistribution of the magnon subbands associated with five distinct phases within GS(0). 
Whereas the BC and CN are sensitive to changes in the magnonic DOS, the INS spectrum is directly sensitive to changes in the number and intensity of magnon
subbands (rather than their distribution) and the magnetization $m$.

A useful check on our results was that the CNs must remain integers for any field and tilt angle $\theta $.  
In our model, the bands exhibit integer CNs (0, $\pm 1$, $\pm 2$, or $\pm 3$) so long as the twist angle $\phi $ is a
multiple of $\pi /3$.  Although the SW frequencies $\omega_n(\vk )$ depend on $\phi $, they remain real for any nonzero $\phi $.  
Hence, the GS remains locally stable for all $\phi $ and the SW frequencies alone cannot be used to determine the physical value(s) for $\phi $.  
While the classical energy (of order $S^2$) is degenerate wrt $\phi $, 
the magnon zero-point energy (of order $S$) is an extremum when $\phi =\pi l/3$ for integer $l$.  We found that $l$ is even below and odd above $H_c'$.
The very shallow minimum of the zero-point energy above $H_c'$ suggests that future work on the temperature dependence of this model would be very interesting.

It would also be interesting to study the emergence of the magnon orbital angular momentum (OAM) with increasing field \cite{Fishman23b}.  Like the BC, the OAM
vanishes in the AF coplanar state with $\theta =0.5\pi$ and will grow as the spins tilt towards the $z$ axis.  Also like the BC, the OAM may show 
signatures of the different phases in GS(0).

Because of the complex behavior revealed in this paper, we conclude that future work on the topological properties of more realistic NC systems must 
be approached cautiously.  Even the simple switching of a magnetic system between up and down states may exhibit rich physics in 
the intermediate NC states.  As we have seen, the topological properties of such systems may be quite intricate.

Admittedly, the model studied in this paper has been based on unrealistic parameters.  We know of no material where the 
DM interaction is so large compared to the exchange interaction and 
easy-axis anisotropy that the spins are confined to the $xy$ plane in zero field.  
Based on Eq.\,(\ref{dc}), the smallest DM interaction $\vert D\vert $ that can confine the spins to the 
$xy$ plane for small anisotropy $K$ is $\sqrt{3}J/3=0.577J$ (with $\vert D\vert =D_c$ and $H_c=0$), which is still probably 3 times larger than found experimentally.
However, the purpose of this paper
was not to describe a real material but rather to demonstrate the challenges that arise when evaluating the topological properties of 
a complex NC spin state with $M>2$ magnetic sublattices.  In that regard, 
this paper has raised many important issues that must be carefully addressed in future calculations for realistic systems.  

This paper has set the stage for the more challenging study of a HC lattice with $\vert D\vert < D_c  $ so that the spins are aligned along $\vz $ in zero field.
A magnetic field along $\vx$ or $\vy $ will then tilt the spins towards the $xy$ plane.  However, even a strong magnetic field would not produce a perfect coplanar state.
Taking $\vert D\vert \approx 0.2 J$, the resulting model would describe the effect of a magnetic field on FM HC materials like CrCl$_3$ \cite{Li22} and CrI$_3$ \cite{McGuire15, Chen18}.
Experiments on the effect of a transverse magnetic field on CrI$_3$ are already underway \cite{Chen21}.  Based on our work,
we expect that a transverse magnetic field will stabilize different GSs and multiple topological phases as the spins bend towards the $xy$ plane.

To conclude, complex NC spins states can exhibit a rich variety of magnetic GSs and topological phases due to the rearrangement
of spin bands within the connected magnonic bands.  Our relatively simple model for a NC spin system exhibits two magnetic GSs and five
topological phases in the low field GS.
We hope that the present calculations will inspire future work on the topological properties of
complex NC spin states. 

Conversations with Peng-Cheng Dai, Gabor Halasz, and David Vanderbilt are gratefully acknowledged.  
Research by R.F. sponsored by the Laboratory
Director's Fund of Oak Ridge National Laboratory.
Research by D.P. supported by the Scientific User Facilities
Division, Office of Basic Energy Sciences, U.S. Department of Energy. 
The data that support the findings of this study are available from the authors
upon reasonable request.

\appendix

\section{Evaluating the BC for NC States}

Any Hamiltonian ${\cal H}$ can be expanded in powers of $1/\sqrt{S}$ as ${\cal H}=E_0+{\cal H}_2+\ldots $, where the GS energy $E_0$ of Eq.\,(\ref{E0}) is
of order $S^2$ and $H_2$ is of order $S$.
We now switch over from the semiclassical Bloch formalism to the SW formalism in order to provide a more precise description of the problem.

In terms of the Bosonic creation and annihilation operators 
$a_{\vk }^{(r)\dagger }$ and $a_{\vk }^{(r)}$ operators, the second-order Hamiltonian ${\cal H}_2$ can be written as
\begin{equation}
{\cal H}_2={\sum_\vk }' {\bf v}_{\vk}^{\dagger }\cdot \underline{{\cal L}}(\vk )\cdot {\bf v}_{\vk },
\label{defL}
\end{equation}
where the vector operators
\begin{equation}
{\bf v}_{\vk } =(a_{\vk }^{(1)},a_{\vk }^{(2)}\ldots  a_{\vk }^{(M)},a_{-\vk}^{(1)\dagger },a_{-\vk }^{(2)\dagger }\ldots a_{-\vk }^{(M)\dagger })
\end{equation}
satisfy 
$[{\bf v}_{\vk },{\bf v}^{\dagger }_{\vk'}] =\underline{N}\,\delta_{\vk ,\vk'}$
with
\begin{equation}
\underline{N} =
\left(
\begin{array}{cc}
\underline{I} &0 \\
0 & -\underline{I} \\
\end{array} \right)
\label{defn}
\end{equation}
and $\underline{I}$ is the $M$-dimensional identity matrix ($M=6$ for the six subband model).   Note that $r=1,\ldots ,M$ are the indices for 
the sites in the magnetic unit cell. 

For each $n$, the vector $\underline{X}^{-1}(\vk )$ obeys the eigenvalue equation \cite{fishmanbook18}
\begin{equation}
\underline{\Lambda} (\vk )\cdot\underline{X}^{-1}(\vk ) = \xi_n(\vk )\,\underline{X}^{-1 }(\vk ),
\label{egv}
\end{equation}
where ${\underline \Lambda}(\vk ) = {\underline N}\cdot \underline{\cal L}(\vk )$
and $\xi_n(\vk )=\hbar \omega_n(\vk )/2$ ($n=0,\ldots ,M$) or $-\hbar \omega_n(-\vk )/2$ ($n=M+1,\ldots ,2M$).
This expression is the quantum analogue of the semiclassical relation 
\begin{equation}
H_2\vert u_n(\vk )\rangle =\hbar \omega_n(\vk )\vert u_n (\vk )\rangle .
\label{bhe}
\end{equation}
Hence, $X^{-1 }(\vk )_{rn}$ can be considered the $n$th eigenfunction of the $2M \times 2M$ magnon energy matrix ${\underline{\Lambda}}(\vk )$.
The physical eigenvalues are given by the positive solutions $\hbar \omega_n(\vk )=2\xi_n(\vk )$ ($n=0,\ldots ,M$)
or, equivalently, by $-2\xi_n(-\vk )$ ($n=M+1,\ldots ,2M$). 

The BC ${\bf \Omega}_n(\vk )$ of Eq.\,(\ref{EqBerry}) 
can be written in terms of the quantum eigenfunctions $\underline{X}^{-1}(\vk )$ as \cite{Fishman23b}
\begin{eqnarray}
&&{\bf \Omega}_n (\vk )= \frac{i}{2\pi}\sum_{r=1}^M\biggl\{ 
\frac{\partial X^{-1}(\vk )_{rn}^*}{\partial \vk } \times \frac{\partial X^{-1}(\vk )_{rn}}{\partial \vk } \nonumber \\
&&- \frac{ \partial X^{-1}(\vk)_{r+M,n}^*}{\partial \vk} \times  \frac {\partial X^{-1}(\vk )_{r+M,n}}{\partial \vk } \biggr\}.
\label{Bfdef}
\end{eqnarray}
For a collinear state, each eigenfunction $X^{-1}(\vk )_{rn}$ can be obtained analytically for every eigenvector $n$ and $r=1,\dots,2M$.
Then, the derivatives wrt $\vk $ can be evaluated without error and the BC can be evaluated analytically.
Examples are provided in Ref. \cite{Fish23}.  

But for a NC spin state, the derivatives $\partial X^{-1}(\vk )_{rn}/\partial \vk $ cannot be evaluated analytically.  The numerical relation 
\begin{equation}
\frac{\partial X^{-1}(\vk )_{rn}}{\partial \vk } = \frac{ X^{-1}(\vk + \vk_0 )_{rn} - X^{-1}(\vk )_{rn} }{\vk_0 }
\end{equation}
for $\vert \vk_0 \vert \ll \vert \vk \vert $ is incomplete because the eigenvalue equation of Eq.\,(\ref{egv}) does not specify the phase of $X^{-1}(\vk )_{rn}$
for a given eigenvector $n$.  When evaluated numerically, $X^{-1}(\vk )_{rn}$ can gain a different random
phase at every wavevector $\vk $ for each $n$.   So for each eigenvector, the derivative defined above can obtain any value.  

However, due to gauge invariance, the overall BC does not depend on the choice of gauge.  A gauge transformation changes the eigenvectors
$X^{-1}(\vk )_{rn}$ by
\begin{equation}
X^{-1}(\vk)_{rn }\rightarrow X^{-1}(\vk )_{rn}\,e^{-i\lambda_n(\vk )},
\label{gtr}
\end{equation}
where $\lambda_n(\vk )$ may depend on $\vk $ and band index $n$ but not on site $r$.  
Under a gauge transformation, 
\begin{equation}
{\bf \Omega}_n(\vk )\rightarrow {\bf \Omega}_n(\vk ),
\end{equation}
which uses the normalization condition
$\underline{X}^{-1}(\vk )\cdot \underline{N} \cdot \underline{X}^{-1\, \dagger }(\vk )=\underline{N}$ or
\begin{equation}
\sum_{r=1}^M \Bigl\{ \vert X^{-1}(\vk )_{rn}\vert^2 - \vert X^{-1}(\vk )_{r+M,n}\vert^2\Bigr\}=1,
\label{sumx}
\end{equation}
which is equivalent to the Bloch normalization condition
$\langle u_n(\vk )\vert u_n(\vk )\rangle =1$.

Thus, we are free to choose an overall phase for each eigenfunction $X^{-1}(\vk )_{rn}$ at a given $n$.  
For every $\vk $, we first order the eigenvalues and eigenvectors from smallest to largest frequencies, $\omega_n(\vk )$ ($n=1,\ldots ,M$)
and $n=M+1,\ldots 2M$.  We then choose $\lambda_n(\vk )$ for each $n$ so that 
\begin{equation}
X^{-1}(\vk)_{M+1,n }\rightarrow X^{-1}(\vk )_{M+1,n}\,e^{-i\lambda_n(\vk )},
\end{equation}
is real \cite{Mnote}.
This choice eliminates the major source of numerical error in the derivatives with respect to $\vk $.

Alternatively, the BC can be evaluated by using semi-classical Bloch functions to transform the derivatives 
$\partial /\partial \vk $ using \cite{Chang96}
\begin{eqnarray}
&&\langle u_{n'}(\vk')\vert \frac{\partial }{\partial \vk }\vert u_n(\vk ) \rangle
=\frac{1}{\epsilon_{n'}(\vk')-\epsilon_n(\vk )}  \nonumber \\
&&\times \langle u_{n'}(\vk ')\vert \frac{\partial \tilde{\cal H} }{ \partial \vk }\vert u_n(\vk ) \rangle ,\\
\,\nonumber 
\end{eqnarray}
where $\tilde{\cal H}=\exp(-i\vk \cdot \vr ){\cal H}\exp (i\vk \cdot \vr )$.  However, this method also confronts 
numerical challenges  \cite{Fuk05, Sesh18} for NC spin states
related to the possibly small denominators in the expression above.

\vfill\eject


\vfill

\begin{thebibliography}{10}

\bibitem{Wang21}
X.~S. Wang and X.~R. Wang.
\newblock {Topological magnonics}.
\newblock {\em Journal of Applied Physics}, 129(15), 04 2021.
\newblock 151101.

\bibitem{Chumak22}
A.~V. Chumak, P.~Kabos, M.~Wu, C.~Abert, C.~Adelmann, A.~O. Adeyeye,
  J.~Åkerman, F.~G. Aliev, A.~Anane, A.~Awad, C.~H. Back, A.~Barman, G.~E.~W.
  Bauer, M.~Becherer, E.~N. Beginin, V.~A. S.~V. Bittencourt, Y.~M. Blanter,
  P.~Bortolotti, I.~Boventer, D.~A. Bozhko, S.~A. Bunyaev, J.~J. Carmiggelt,
  R.~R. Cheenikundil, F.~Ciubotaru, S.~Cotofana, G.~Csaba, O.~V. Dobrovolskiy,
  C.~Dubs, M.~Elyasi, K.~G. Fripp, H.~Fulara, I.~A. Golovchanskiy,
  C.~Gonzalez-Ballestero, P.~Graczyk, D.~Grundler, P.~Gruszecki, G.~Gubbiotti,
  K.~Guslienko, A.~Haldar, S.~Hamdioui, R.~Hertel, B.~Hillebrands, T.~Hioki,
  A.~Houshang, C.-M. Hu, H.~Huebl, M.~Huth, E.~Iacocca, M.~B. Jungfleisch,
  G.~N. Kakazei, A.~Khitun, R.~Khymyn, T.~Kikkawa, M.~Kläui, O.~Klein, J.~W.
  Kłos, S.~Knauer, S.~Koraltan, M.~Kostylev, M.~Krawczyk, I.~N. Krivorotov,
  V.~V. Kruglyak, D.~Lachance-Quirion, S.~Ladak, R.~Lebrun, Y.~Li, M.~Lindner,
  R.~Macêdo, S.~Mayr, G.~A. Melkov, S.~Mieszczak, Y.~Nakamura, H.~T. Nembach,
  A.~A. Nikitin, S.~A. Nikitov, V.~Novosad, J.~A. Otálora, Y.~Otani, A.~Papp,
  B.~Pigeau, P.~Pirro, W.~Porod, F.~Porrati, H.~Qin, B.~Rana, T.~Reimann,
  F.~Riente, O.~Romero-Isart, A.~Ross, A.~V. Sadovnikov, A.~R. Safin,
  E.~Saitoh, G.~Schmidt, H.~Schultheiss, K.~Schultheiss, A.~A. Serga,
  S.~Sharma, J.~M. Shaw, D.~Suess, O.~Surzhenko, K.~Szulc, T.~Taniguchi,
  M.~Urbánek, K.~Usami, A.~B. Ustinov, T.~van~der Sar, S.~van Dijken, V.~I.
  Vasyuchka, R.~Verba, S.~Viola Kusminskiy, Q.~Wang, M.~Weides, M.~Weiler,
  S.~Wintz, S.~P. Wolski, and X.~Zhang.
\newblock Advances in magnetics roadmap on spin-wave computing.
\newblock {\em IEEE Transactions on Magnetics}, 58(6):1--72, 2022.

\bibitem{Sheka22}
Denis~D. Sheka, Oleksandr~V. Pylypovskyi, Oleksii~M. Volkov, Kostiantyn~V.
  Yershov, Volodymyr~P. Kravchuk, and Denys Makarov.
\newblock Fundamentals of curvilinear ferromagnetism: Statics and dynamics of
  geometrically curved wires and narrow ribbons.
\newblock {\em Small}, 18(12):2105219, 2022.

\bibitem{Onose2010}
Y.~Onose, T.~Ideue, H.~Katsura, Y.~Shiomi, N.~Nagaosa, and Y.~Tokura.
\newblock Observation of the magnon {H}all effect.
\newblock {\em Science}, 329(5989):297--299, 2010.

\bibitem{Ideue12}
T.~Ideue, Y.~Onose, H.~Katsura, Y.~Shiomi, S.~Ishiwata, N.~Nagaosa, and
  Y.~Tokura.
\newblock Effect of lattice geometry on magnon {H}all effect in ferromagnetic
  insulators.
\newblock {\em Phys. Rev. B}, 85:134411, Apr 2012.

\bibitem{Hirschberger15a}
Max Hirschberger, Jason~W. Krizan, R.~J. Cava, and N.~P. Ong.
\newblock Large thermal {H}all conductivity of neutral spin excitations in a
  frustrated quantum magnet.
\newblock {\em Science}, 348(6230):106--109, 2015.

\bibitem{Hirschberger15b}
Max Hirschberger, Robin Chisnell, Young~S. Lee, and N.~P. Ong.
\newblock Thermal {H}all effect of spin excitations in a {K}agome magnet.
\newblock {\em Phys. Rev. Lett.}, 115:106603, Sep 2015.

\bibitem{Murakami17}
Shuichi Murakami and Akihiro Okamoto.
\newblock Thermal {H}all effect of magnons.
\newblock {\em Journal of the Physical Society of Japan}, 86(1):011010, 2017.

\bibitem{Uchida10}
Ken-ichi Uchida, Hiroto Adachi, Takeru Ota, Hiroyasu Nakayama, Sadamichi
  Maekawa, and Eiji Saitoh.
\newblock Observation of longitudinal spin-{S}eebeck effect in magnetic
  insulators.
\newblock {\em Applied Physics Letters}, 97(17):172505, 2010.

\bibitem{Wu16}
Stephen~M. Wu, Wei Zhang, Amit KC, Pavel Borisov, John~E. Pearson, J.~Samuel
  Jiang, David Lederman, Axel Hoffmann, and Anand Bhattacharya.
\newblock Antiferromagnetic spin {S}eebeck effect.
\newblock {\em Phys. Rev. Lett.}, 116:097204, Mar 2016.

\bibitem{Buttner2000}
O.~Büttner, M.~Bauer, A.~Rueff, S.O. Demokritov, B.~Hillebrands, A.N. Slavin,
  M.P. Kostylev, and B.A. Kalinikos.
\newblock Space- and time-resolved {B}rillouin light scattering from nonlinear
  spin-wave packets.
\newblock {\em Ultrasonics}, 38(1):443--449, 2000.

\bibitem{Katsura2010}
Hosho Katsura, Naoto Nagaosa, and Patrick~A. Lee.
\newblock Theory of the thermal {H}all effect in quantum magnets.
\newblock {\em Phys. Rev. Lett.}, 104:066403, Feb 2010.

\bibitem{Mat11a}
Ryo Matsumoto and Shuichi Murakami.
\newblock Theoretical prediction of a rotating magnon wave packet in
  ferromagnets.
\newblock {\em Phys. Rev. Lett.}, 106:197202, May 2011.

\bibitem{Mat11b}
Ryo Matsumoto and Shuichi Murakami.
\newblock Rotational motion of magnons and the thermal {H}all effect.
\newblock {\em Phys. Rev. B}, 84:184406, Nov 2011.

\bibitem{Mong11}
Roger S.~K. Mong and Vasudha Shivamoggi.
\newblock Edge states and the bulk-boundary correspondence in {D}irac
  {H}amiltonians.
\newblock {\em Phys. Rev. B}, 83:125109, Mar 2011.

\bibitem{Zhang13}
Lifa Zhang, Jie Ren, Jian-Sheng Wang, and Baowen Li.
\newblock Topological magnon insulator in insulating ferromagnet.
\newblock {\em Phys. Rev. B}, 87:144101, Apr 2013.

\bibitem{Mook14a}
Alexander Mook, J\"urgen Henk, and Ingrid Mertig.
\newblock Magnon {H}all effect and topology in {K}agome lattices: A theoretical
  investigation.
\newblock {\em Phys. Rev. B}, 89:134409, Apr 2014.

\bibitem{Shindou13}
Ryuichi Shindou, Ryo Matsumoto, Shuichi Murakami, and Jun-ichiro Ohe.
\newblock Topological chiral magnonic edge mode in a magnonic crystal.
\newblock {\em Phys. Rev. B}, 87:174427, May 2013.

\bibitem{Owerre16}
S.A. Owerre.
\newblock {Topological honeycomb magnon Hall effect: A calculation of thermal
  Hall conductivity of magnetic spin excitations}.
\newblock {\em Journal of Applied Physics}, 120(4):043903, 07 2016.

\bibitem{Owerre16b}
S.~A. Owerre.
\newblock Magnon {H}all effect in {AB}-stacked bilayer honeycomb quantum
  magnets.
\newblock {\em Phys. Rev. B}, 94:094405, Sep 2016.

\bibitem{Mook21}
Alexander Mook, Kirill Plekhanov, Jelena Klinovaja, and Daniel Loss.
\newblock Interaction-stabilized topological magnon insulator in ferromagnets.
\newblock {\em Phys. Rev. X}, 11:021061, Jun 2021.

\bibitem{Li21b}
Yu-Hang Li and Ran Cheng.
\newblock Magnonic {S}u-{S}chrieffer-{H}eeger model in honeycomb ferromagnets.
\newblock {\em Phys. Rev. B}, 103:014407, Jan 2021.

\bibitem{Lu21}
Yu-Shan Lu, Jian-Lin Li, and Chien-Te Wu.
\newblock Topological phase transitions of {D}irac magnons in honeycomb
  ferromagnets.
\newblock {\em Phys. Rev. Lett.}, 127:217202, Nov 2021.

\bibitem{Fishman23}
Randy~S. Fishman, Lucas Lindsay, and Satoshi Okamoto.
\newblock Exact results for the orbital angular momentum of magnons on
  honeycomb lattices.
\newblock {\em Journal of Physics: Condensed Matter}, 35(1):015801, nov 2022.

\bibitem{Zyuzin16}
Vladimir~A. Zyuzin and Alexey~A. Kovalev.
\newblock Magnon spin {N}ernst effect in antiferromagnets.
\newblock {\em Phys. Rev. Lett.}, 117:217203, Nov 2016.

\bibitem{Cheng16}
Ran Cheng, Satoshi Okamoto, and Di~Xiao.
\newblock Spin {N}ernst effect of magnons in collinear antiferromagnets.
\newblock {\em Phys. Rev. Lett.}, 117:217202, Nov 2016.

\bibitem{Nakata17}
Kouki Nakata, Se~Kwon Kim, Jelena Klinovaja, and Daniel Loss.
\newblock Magnonic topological insulators in antiferromagnets.
\newblock {\em Phys. Rev. B}, 96:224414, Dec 2017.

\bibitem{Kawano19}
Masataka Kawano and Chisa Hotta.
\newblock Thermal hall effect and topological edge states in a square-lattice
  antiferromagnet.
\newblock {\em Phys. Rev. B}, 99:054422, Feb 2019.

\bibitem{Neumann22}
Robin~R. Neumann, Alexander Mook, J\"urgen Henk, and Ingrid Mertig.
\newblock Thermal {H}all effect of magnons in collinear antiferromagnetic
  insulators: Signatures of magnetic and topological phase transitions.
\newblock {\em Phys. Rev. Lett.}, 128:117201, Mar 2022.

\bibitem{Go24}
Gyungchoon Go, Daehyeon An, Hyun-Woo Lee, and Se~Kwon Kim.
\newblock Magnon orbital {N}ernst effect in honeycomb antiferromagnets without
  spin–orbit coupling.
\newblock {\em Nano Letters}, 24(20):5968--5974, 2024.
\newblock PMID: 38682941.

\bibitem{Sticlet12}
Doru Sticlet, Frederic Pi\'echon, Jean-No\"el Fuchs, Pavel Kalugin, and Pascal
  Simon.
\newblock Geometrical engineering of a two-band {C}hern insulator in two
  dimensions with arbitrary topological index.
\newblock {\em Phys. Rev. B}, 85:165456, Apr 2012.

\bibitem{Fujiwara22}
Kosuke Fujiwara, Sota Kitamura, and Takahiro Morimoto.
\newblock Thermal {H}all responses in frustrated honeycomb spin systems.
\newblock {\em Phys. Rev. B}, 106:035113, Jul 2022.

\bibitem{Chen23}
Qi-Hui Chen, Fei-Jie Huang, and Yong-Ping Fu.
\newblock Damped topological magnons in honeycomb antiferromagnets.
\newblock {\em Phys. Rev. B}, 108:024409, Jul 2023.

\bibitem{Haldane88}
F.~D.~M. Haldane.
\newblock Model for a quantum hall effect without {L}andau levels:
  Condensed-matter realization of the ``parity anomaly".
\newblock {\em Phys. Rev. Lett.}, 61:2015--2018, Oct 1988.

\bibitem{Wang23}
Xinhao Wang, Mohammad~Tomal Hossain, T.~R. Thapaliya, Durga Khadka, Sergi
  Lendinez, Hang Chen, Matthew~F. Doty, M.~Benjamin Jungfleisch, S.~X. Huang,
  Xin Fan, and John~Q. Xiao.
\newblock Spin currents with unusual spin orientations in noncollinear {W}eyl
  antiferromagnetic {M}n$_{3}${S}n.
\newblock {\em Phys. Rev. Mater.}, 7:034404, Mar 2023.

\bibitem{Martini23}
Mickey Martini, Helena Reichlova, Laura~T. Corredor, Dominik Kriegner, Yejin
  Lee, Luca Tomarchio, Kornelius Nielsch, Ali~G. Moghaddam, Jeroen van~den
  Brink, Bernd B\"uchner, Sabine Wurmehl, Vitaliy Romaka, and Andy Thomas.
\newblock Anomalous hall effect and magnetoresistance in microribbons of the
  magnetic {W}eyl semimetal candidate {P}r{R}h{C}$_{2}$.
\newblock {\em Phys. Rev. Mater.}, 7:104205, Oct 2023.

\bibitem{Li23}
Z.~Li, T.~Chirac, J.~Tranchida, V.~Garcia, S.~Fusil, V.~Jacques, J.-Y.
  Chauleau, and M.~Viret.
\newblock Multiferroic {S}kyrmions in {B}i{F}e{O}$_{3}$.
\newblock {\em Phys. Rev. Res.}, 5:043109, Nov 2023.

\bibitem{Soenen23}
Maarten Soenen and Milorad~V. Milo\ifmmode \check{s}\else
  \v{s}\fi{}evi\ifmmode~\acute{c}\else \'{c}\fi{}.
\newblock Tunable magnon topology in monolayer {C}r{I}$_{3}$ under external
  stimuli.
\newblock {\em Phys. Rev. Mater.}, 7:084402, Aug 2023.

\bibitem{Hayami23}
Satoru Hayami and Hiroaki Kusunose.
\newblock Time-reversal switching responses in antiferromagnets.
\newblock {\em Phys. Rev. B}, 108:L140409, Oct 2023.

\bibitem{Fernandes24}
Rafael~M. Fernandes, Vanuildo~S. de~Carvalho, Turan Birol, and Rodrigo~G.
  Pereira.
\newblock Topological transition from nodal to nodeless zeeman splitting in
  altermagnets.
\newblock {\em Phys. Rev. B}, 109:024404, Jan 2024.

\bibitem{Guo24}
Yaqin Guo, Jing Zhang, Purnima~P. Balakrishnan, Alexander~J. Grutter, Baishun
  Yang, Michael~R. Fitzsimmons, Timothy~R. Charlton, Haile Ambaye, Xu~Zhang,
  Hanshen Huang, Zhi Huang, Jinyan Chen, Chenyang Guo, Xiufeng Han, Kang~L.
  Wang, and Hao Wu.
\newblock Controllable conical magnetic structure and spin-orbit-torque
  switching in symmetry-broken ferrimagnetic films.
\newblock {\em Phys. Rev. Appl.}, 21:014045, Jan 2024.

\bibitem{Jana24}
Apu~Kumar Jana, Kyung~Jae Lee, Sanghoon Lee, Xinyu Liu, Margaret Dobrowolska,
  and Jacek~K. Furdyna.
\newblock Magnetization switching by spin-orbit torque in crystalline
  ({G}a,{M}n)({A}s,{P}) film deposited on a vicinal {G}a{A}s substrate.
\newblock {\em Phys. Rev. B}, 110:054422, Aug 2024.

\bibitem{fishmanbook18}
Randy~S. Fishman, Jaime Fernandez-Baca, and Toomas R{\~o}{\~o}m.
\newblock {\em Spin-Wave Theory and its Applications to Neutron Scattering and
  THz Spectroscopy}.
\newblock Morgan and Claypool Publishers, San Rafael, 2018.

\bibitem{Kubo52}
Ryogo Kubo.
\newblock The spin-wave theory of antiferromagnetics.
\newblock {\em Phys. Rev.}, 87:568--580, Aug 1952.

\bibitem{McClarty14}
Paul~A. McClarty, Pawel Stasiak, and Michel J.~P. Gingras.
\newblock Order-by-disorder in the {XY} pyrochlore antiferromagnet.
\newblock {\em Phys. Rev. B}, 89:024425, Jan 2014.

\bibitem{Vanderbilt18}
David Vanderbilt.
\newblock {\em Berry Phases in Electronic Structure Theory: Electric
  Polarization, Orbital Magnetization and Topological Insulators}.
\newblock Cambridge University Press, Cambridge, 2018.

\bibitem{Fishman23b}
Randy~S. Fishman.
\newblock Gauge-invariant measure of the magnon orbital angular momentum.
\newblock {\em Phys. Rev. B}, 107:214434, Jun 2023.

\bibitem{Li22}
Xun Li, Seung-Hwan Do, Jiaqiang Yan, Michael~A. McGuire, Garrett~E. Granroth,
  Sai Mu, Tom Berlijn, Valentino~R. Cooper, Andrew~D. Christianson, and Lucas
  Lindsay.
\newblock Phonons and phase symmetries in bulk {C}r{C}l$_3$ from scattering
  measurements and theory.
\newblock {\em Acta Materialia}, 241:118390, 2022.

\bibitem{McGuire15}
Michael~A. McGuire, Hemant Dixit, Valentino~R. Cooper, and Brian~C. Sales.
\newblock Coupling of crystal structure and magnetism in the layered,
  ferromagnetic insulator {C}r{I}$_3$.
\newblock {\em Chemistry of Materials}, 27(2):612--620, 2015.

\bibitem{Chen18}
Lebing Chen, Jae-Ho Chung, Bin Gao, Tong Chen, Matthew~B. Stone, Alexander~I.
  Kolesnikov, Qingzhen Huang, and Pengcheng Dai.
\newblock Topological spin excitations in honeycomb ferromagnet {C}r{I}$_3$.
\newblock {\em Phys. Rev. X}, 8:041028, Nov 2018.

\bibitem{Chen21}
Lebing Chen, Jae-Ho Chung, Matthew~B. Stone, Alexander~I. Kolesnikov, Barry
  Winn, V.~Ovidiu Garlea, Douglas~L. Abernathy, Bin Gao, Mathias Augustin,
  Elton J.~G. Santos, and Pengcheng Dai.
\newblock Magnetic field effect on topological spin excitations in
  {C}r{I}$_{3}$.
\newblock {\em Phys. Rev. X}, 11:031047, Aug 2021.

\bibitem{Fish23}
Randy~S. Fishman, Tom Berlijn, Jack Villanova, and Lucas Lindsay.
\newblock Magnon orbital angular momentum of ferromagnetic honeycomb and zigzag
  lattice models.
\newblock {\em Phys. Rev. B}, 108:214402, Dec 2023.

\bibitem{Mnote}
Any component of the eigenvector can be chosen to be real so long as it has a
  significant amplitude.

\bibitem{Chang96}
Ming-Che Chang and Qian Niu.
\newblock Berry phase, hyperorbits, and the {H}ofstadter spectrum:
  Semiclassical dynamics in magnetic {B}loch bands.
\newblock {\em Phys. Rev. B}, 53:7010--7023, Mar 1996.

\bibitem{Fuk05}
Takahiro Fukui, Yasuhiro Hatsugai, and Hiroshi Suzuki.
\newblock Chern numbers in discretized {B}rillouin zone: Efficient method of
  computing (spin) {H}all conductances.
\newblock {\em Journal of the Physical Society of Japan}, 74(6):1674--1677,
  2005.

\bibitem{Sesh18}
Ranjani Seshadri and Diptiman Sen.
\newblock Topological magnons in a {K}agome-lattice spin system with {XXZ} and
  {D}zyaloshinskii-{M}oriya interactions.
\newblock {\em Phys. Rev. B}, 97:134411, Apr 2018.

\end{thebibliography}
\end{document}